\documentclass[article]{JHEP3}

\usepackage{amsmath,epsfig}

\usepackage{amssymb,amsfonts}

\title{Strings in an electric field, and the Milne Universe}
\preprint{\hepth{0307280}\\LPTHE-03-21\\WIS/20/03-JUL-DPP}

\author{M. Berkooz\footnote{Incumbent of the Recanati career
development chair for energy research}\\
Weizmann Institute of Science,\\ Rehovot 76100, Israel\\{\tt E-mail:
berkooz@wisemail.weizmann.ac.il}}

\author{B. Pioline\\
LPTHE, Universit\'es Paris 6 et 7, 4 place Jussieu, \\
75252 Paris cedex 05, France\\{\tt E-mail:
pioline@lpthe.jussieu.fr}}

\abstract{
Arguably the simplest model of a cosmological singularity in string theory,
the Lorentzian orbifold $\Real^{1,1}/boost$
is known to lead to severe divergences
in scattering amplitudes of untwisted states, indicating a large backreaction
toward the singularity. In this work we take a first step in investigating
whether condensation of twisted states may remedy this problem and resolve
the spacelike singularity. By using the formal analogy with charged open
strings in an electric field, we argue that, contrary to earlier claims,
twisted sectors do contain physical scattering states, which can be viewed
as charged particles in an electric field. Correlated pairs of twisted
states will therefore be produced, by the ordinary Schwinger mechanism.
For open strings in an electric field, on-shell wave functions
for the zero-modes are determined, and shown
to analytically 
continue to non-normalizable modes of the usual Landau harmonic oscillator
in Euclidean space. Closed strings scattering states of the Milne orbifold
continue to non-normalizable modes in an unusual Euclidean orbifold
of $\Real^2$ by a rotation by an irrational angle. 
Irrespective of the formal analogy with the Milne
Universe, open strings in a constant electric field, or colliding D-branes,
may also serve as a useful laboratory to study time-dependence in
string theory.}

\makeatletter
\renewcommand{\subsubsection}{\@startsection{subsubsection}{3}{0mm}{-\baselineskip}{0.5\baselineskip}{\normalfont\normalsize\it}}
\makeatother

\newcommand{\pa}{\partial}

\newcommand{\p}{\partial}

\newcommand{\nn}{\nonumber}
\newcommand{\eps}{\epsilon}
\newcommand{\Real}{\mathbb{R}}
\newcommand{\Zint}{\mathbb{Z}}

\newcommand{\sgn}{\mbox{sgn}}

\def\bea{\begin{eqnarray}}
\def\eea{\end{eqnarray}}
\def\be{\begin{equation}}
\def\ee{\end{equation}}
\def\ba{\begin{align}}
\def\ea{\end{align}}
\def\bse{\begin{subequations}}
\def\ese{\end{subequations}}
\def\bi{\begin{itemize}}
\def\ei{\end{itemize}}

\def\a{\alpha}
\def\talpha{\tilde\alpha}
\def\t{\theta}

\def\ta{\tilde \alpha}
\def\1F1{{}_1\!F_1}
\def\2F0{{}_2\!F_0}
\def\ie{{\it i.e.} }

\begin{document}
\maketitle
\setcounter{tocdepth}{2}
\tableofcontents

\section{Introduction}

One of the most fundamental problems in quantum gravity
is the nature of the Big Bang singularity, or more general
space-like singularities. This is interesting both from a theoretical
viewpoint, as it is likely to cause a breakdown of effective field theory,
and from a practical viewpoint, as our Universe may have started in such a
singularity. In the context of string theory, this issue is also
tied with the problem of time dependence, and might play a role
in solving the still mysterious vacuum selection problem.

A possible avenue into the study of gravitational
singularities is to investigate
exact two-dimensional conformal field theories
which realize them in target space. A particularly simple
example is the orbifold of the two-dimensional Minkowski space $\Real^{1,1}$ by
a discrete boost $X^\pm \equiv e^{\pm \beta} X^\pm$,
tensored with a $\hat c=8$ superconformal field theory
such as flat Euclidean space $\Real^8$ \cite{Horowitz:ap,Khoury:2001bz}.
Part of the geometry described
by this orbifold CFT corresponds to the Milne Universe, \ie
a circle shrinking linearly in time till a Big Crunch,
or growing linearly in time after a Big Bang
(see Figure \ref{milneorb}). It also includes two
disconnected regions with closed time-like curves (CTC) attached
to the cosmological singularity, whose existence may yet
be a fatal flaw or a bounty. Despite being a very non-generic case of
a Kasner singularity, the same geometry arises locally in many
other examples based on gauged Wess-Zumino models \cite{Nappi:1992kv,
Elitzur:2002rt,Craps:2002ii,Elitzur:2002vw}. Other time-dependent
orbifolds have been discussed recently in \cite{lms,Balasubramanian:2002ry,
Simon:2002ma,Russo:2003ky}.

\FIGURE{\label{milneorb}
\hfill\epsfig{file=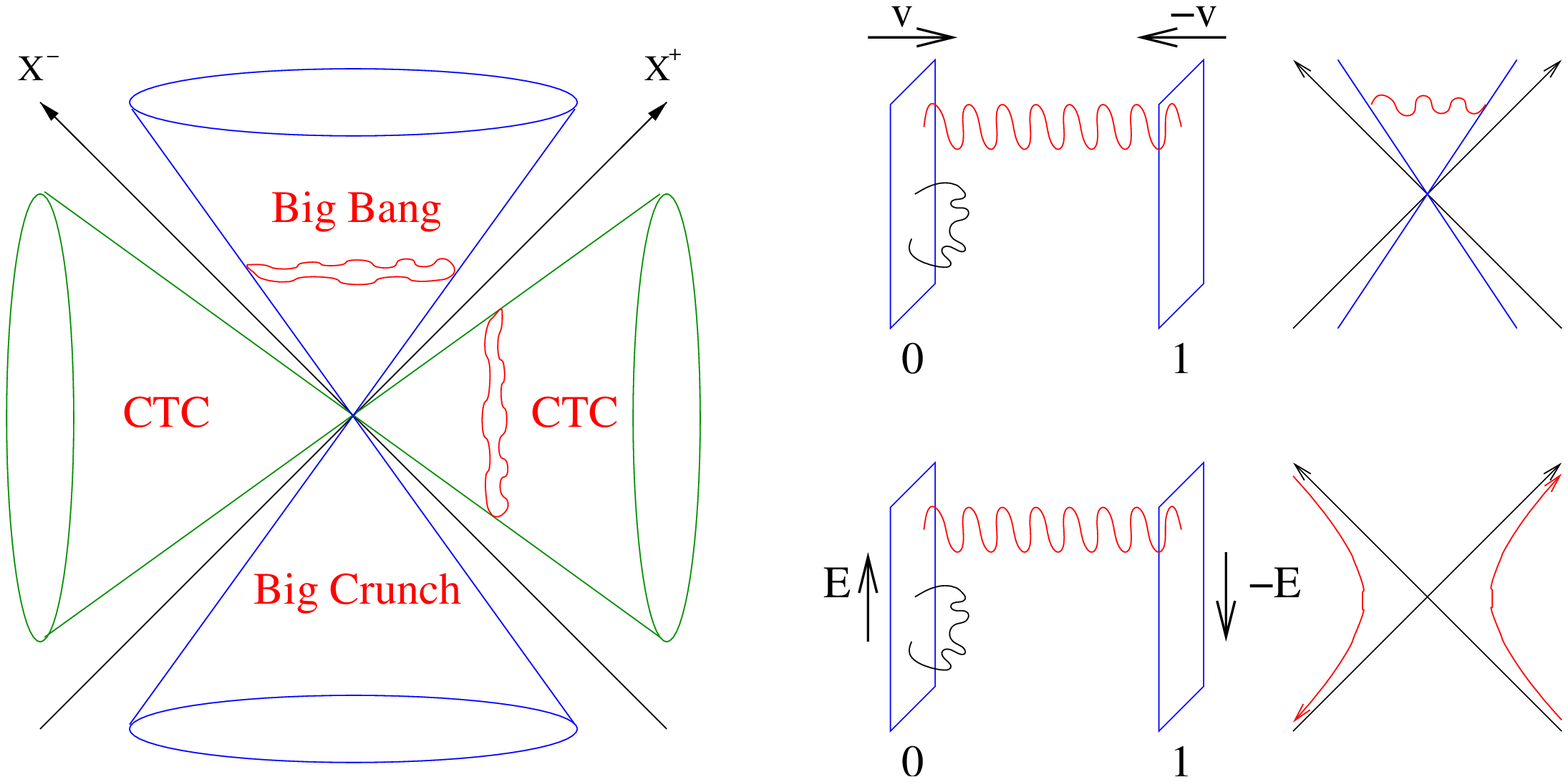,height=6cm}\hfill
\caption{Milne orbifold, colliding D-branes and open strings in an
electric field.}}

Although the Lorentzian orbifold CFT
may seem to be well defined using ordinary orbifold
techniques, tree-level scattering amplitudes of untwisted string states
are divergent, casting doubt on the validity of
perturbation theory \cite{Seiberg:2002hr,Berkooz:2002je} 
(see \cite{lms,Fabinger:2002kr}
for related work in the case of null singularities).
The divergences may be traced back
to large graviton exchange near the singularity \cite{Berkooz:2002je},
where incoming particles become infinitely blue shifted.
A non-perturbative instability toward large black hole creation has
also been argued to result from the same mechanism \cite{Horowitz:2002mw}.
On the other hand, the dynamics in the twisted sector is
far less understood, and leaves open the possibility
that a condensation of twisted states would remedy the afore-mentioned
difficulties. This expectation is supported by the often noticed formal
similarity, at the level of the CFT,
of this closed string background with open strings in an electric
field (see \cite{bh} for further time-dependent open/closed string analogies).
Indeed, the structure of excited modes of first quantized
closed strings in the {\it twisted} sector of the Milne orbifold essentially
amounts to two copies of that of {\it charged} open strings in a
constant electric field, while the zero-modes are identical
between the two cases without doubling, just as in flat space.
As we shall see, in both the electric and Milne case,
the zero-mode degrees of freedom describe the motion of a
charged particle (the string center of motion) in an electric field.

To be more precise, consider two parallel D1-branes
with anti-parallel Abelian electric fields $F^{(0)}=e dx^+ dx^-\ ,
F^{(1)}=-e dx^+ dx^-$ (see Figure \ref{milneorb}).
Open strings stretching between the two carry a
net electric charge under the electric field $F^{(0)}-F^{(1)}$,
and are analogous to closed strings in the $w$-twisted sector of the orbifold
with boost parameter $\beta$, upon identifying $2\pi e \sim - w \beta$
for small $e$. Open strings in the 0-0 or 1-1 sectors on the other hand
have no total charge, and as we will see, 
behave analogously to closed strings in the
untwisted sector. In contrast to the physics of closed strings
near a cosmological singularity,
the dynamics of charged open strings in an electric field is fairly well
understood \cite{Fradkin:1985qd,
Abouelsaood:gd,Burgess:1986dw,Bachas:bh}: by the
usual Schwinger mechanism \cite{Schwinger:nm}, pairs of charged open strings
are created from the vacuum, and cause the electric field to decay
in a sequence of plasma oscillations,
as they move off to infinity and discharge the condenser plates that
created the electric field in the first
place \cite{Cooper:kf,Kluger:1991ib,Tomaras:2000ag,Tomaras:2001vs}.
At late times, one therefore
recovers a situation with zero electric field.

As a way to make contact with the Milne Universe, one
may consider the T-dual of the electric field configuration,
namely two D-branes moving with opposite velocities $\pm e$, respectively:
for $e\neq 0$, the distance between the two D-branes decreases
linearly with time until they collide, just like the radius of
the circle in the Milne universe before the Big Crunch.
Just as in the electric field case, stretched open strings are pair-produced
as the D-branes move away from (resp. toward) each other, and decelerate
(resp. accelerate) the motion \cite{Bachas:1995kx,Douglas:1996yp}.
In contrast to the electric field case, the final outcome
of the D-brane collision is however less clear, as it depends on the
ratio between production and recombination rate of open strings,
as well as on the details of the bound state formation process.

By analogy with these open string processes, one may therefore expect that
the creation of pairs of twisted closed strings
will slow down the contraction
rate of the Milne Universe, and possibly prevent the compact circle
from reaching zero-size.
Indeed, once produced, the twisted closed strings 
contribute a tensive energy that grows linearly with the radius of the Milne
universe, thus mimicking the effect of a two-dimensional positive cosmological
constant\footnote{We thank B.Craps for a discussion on this point.}.  
Whether the circle will further re-expand indefinitely
or reach a constant radius cannot be decided on the basis
of this analogy alone, as the D-brane collision process above illustrates.
In contrast to the resolution of the usual time-like orbifold
singularities by condensation of a coherent state of the twisted sector
moduli field, this process involves the condensation of correlated
multi-particle states, which go beyond standard
string perturbation techniques. We hope to return to this problem 
in a future publication.

More modestly, our aim in this paper will be to understand the
kinematics of twisted sector states, for which purpose
the analogy with open strings in an electric field will turn out
to be quite illuminating. In particular, the implicit assumption
we made above that twisted sectors contain physical
states appears to contradict a recent investigation of the
one-loop amplitude in the Lorentzian orbifold  \cite{Nekrasov:2002kf}.
As the analogy with open strings
in electric fields will make it clear,
upon appropriate quantization of the string zero-modes there 
does in fact exist a continuum of delta-normalizable physical 
scattering states,
describing the unbounded trajectories of charged particle in  an
electric field.
Such a spectrum in fact consistent with the one-loop amplitude computed
in \cite{Nekrasov:2002kf,Cornalba:2002fi}, as the latter
may be viewed either as the contribution of discrete Euclidean
or continuous Lorentzian states.

The physical origin of this continuous spectrum of physical states
is clear: in either  the electric field and
Milne universe cases, these states arise
from quantizing the charged Klein-Gordon equation in two dimensions.
In static coordinates, they are the scattering states of the inverted harmonic
oscillator. In light-cone coordinates,
they are eigenmodes of the scaling operator
on phase space\footnote{The latter quantization scheme
was in fact suggested in a footnote of \cite{Nekrasov:2002kf}.} $H=pq$.
In Rindler coordinates, best suited for the
Lorentzian orbifold case where one needs to project on states
with integer boost momentum $J$, they correspond to
eigenstates of the Schr\"odinger equation with Liouville-like potential
$V(y)=M^2 e^{2y} - (J+ \frac12 \nu e^{2y})^2$.
In all these cases, the potential is unbounded from below, so that there is
no vacuum in the first quantized approach. The second quantized theory
however  is  well defined and has a variety of {\it in} and {\it out} states,
due to Schwinger pair creation (and,
in the Rindler case, Hawking particle production). The  latter  can
be  understood very simply as the tunneling under the 
barrier in these unbounded potentials  
\cite{Casher:wy,Brezin:xf,Parikh:1999mf}.

Having found that the charged (resp. twisted) sectors of the open
(resp. closed) string contain physical states, an intermediate goal
before taking their back-reaction into account is to determine their
scattering amplitudes, and check whether they turn out to be less singular
than those of untwisted states. For this one needs to perform
an analytic Wick rotation both on the world-sheet and in target space,
an often perilous task in generic time-dependent backgrounds.
Despite the lack of a global time-like Killing vector, we show that
the Lorentzian orbifold does admit a well defined analytic continuation,
upon simultaneously Wick rotating the target space light-cone
coordinates $X^\pm$ into a pair of complex conjugate coordinates $Z,\bar Z$,
as well as continuing the real boost parameter $\beta$ to an
imaginary value.  
In the open string case, this prescription is precisely what allows
one to convert an electric field in Minkowski space to a magnetic field in
Euclidean space. 
In the closed string case, one obtains instead an orbifold of the Euclidean
plane by a rotation of angle $\beta$. Instead of reducing to the usual
rotation orbifold $\mathbb{C}/\Zint_N$ for rational values of $\beta$,
as often assumed in other instances of conical singularities
\cite{Dabholkar:1994ai,Lowe:1994ah}, we find that it behaves continuously
with respect to the rotation angle $\beta$, thanks to the inclusion of
string states that wind several times around the origin. This is
of course necessary if the analytic continuation back to the Lorentzian
orbifold is to make sense. In fact, the same rotation orbifold construction
arises in the context of closed strings in plane gravitational waves
supported by a Neveu-Schwarz magnetic flux \cite{Kiritsis:1994ij,
Kiritsis:2002kz},
for which usual current algebra techniques allow one  to compute the scattering
amplitudes \cite{D'Appollonio:2003dr}.

Although this analytic continuation of the electric field (resp. Lorentzian
orbifold) gives the same result as the continuation of a magnetic field
(resp. Euclidean rotation orbifold) with respect to a transverse time
coordinate, it is important to realize that the electric/Milne
physical states continue to an altogether different set of observables.
While the states usually considered in a magnetic field are the usual
normalizable Landau states with positive energy, the analytic
continuation of the scattering states in an electric field involves instead
negative energy non-normalizable states of the harmonic oscillator;
these states blow up in the Euclidean direction where the particle comes from.
Similarly, the states of interest in the rotation orbifold case are not
the usual localized twisted states, but non-normalizable states which 
blow up either at the origin of the plane
or at infinity. It would be interesting
to adapt the techniques in \cite{D'Appollonio:2003dr} for such states,
although we will not attempt this in this paper.

The outline of this paper is as follows. In Section 2, we review
some of the literature on the first quantization of
charged open strings in a constant electric field, and of
closed strings in the twisted sector of the Milne
orbifold. In Section 3, we propose a quantization scheme
for the zero-mode sector which leads to  physical states in the
twisted sectors. We explain its physical origin in static coordinates,
which are most convenient to discuss Schwinger pair creation in
an electric field,
and illustrate the prescription for excited states up to level 1.
In Section 4 we turn to Milne space, which requires 
quantizing the zero-mode wave functions from the point of view of
an accelerated observer, \ie in Rindler coordinates. We evaluate
semi-classically the distribution of produced pairs, noting that
it can be obtained by projecting the usual homogeneous pair production
in an electric field to boost invariant states. Section 5 is devoted
to analyzing the continuation to Euclidean space, both in the electric
field and Lorentzian orbifold cases. Our conclusions are presented
in Section 6, together with comments on open issues.
Appendices contain further material on light-cone quantization,
Wick rotation and a review of parabolic cylinder and Whittaker functions.

\section{First quantization, reviewed}

In this section, we first review elementary aspects
of charged open strings in an electric field,
following \cite{Bachas:bh} up to small changes
in notations. We then turn to twisted closed strings in the Lorentzian
orbifold, following \cite{Nekrasov:2002kf}, and discuss how the two
problems are related.

\subsection{Charged particle in an electric field}
Before turning to the case of open strings in a constant electric field,
it is useful to recall some basic features of charged particles in an
electric field.

\subsubsection{Classical trajectories and conserved charges}

The Lagrangian for a particle of mass $m$ and
unit charge in a constant electric field $F=e~dx^+ \wedge dx^-$ reads
\be
\label{lpa}
L=\frac12 m \left( - 2 \pa_\tau X^+ \pa_\tau X^-
+ (\pa_\tau X^i)^2 \right) + \frac{e}{2} \left( X^+ \pa_\tau X^- -
X^- \pa_\tau X^+ \right)
\ee
where we defined the light-cone coordinates $X^\pm=(X^0\pm X^1)/\sqrt{2}$,
and work with the mostly plus metric $ds^2=-2 dX^+ dX^- + (dX^i)^2$.
The equations of motion are easily integrated to
\be
\label{traj}
X^\pm = x_0^\pm \pm \frac{a_0^\pm}{e} e^{\pm e \tau/m}\ ,\quad
X^i= x_0^i + \frac{p^i}{m} \tau
\ee
hence classical trajectories are hyperbolas centered at
an arbitrary point $(x_0^+,x_0^-)$ and asymptoting to 
the light-cone (see Figure 2). The canonical momenta
\be
\pi^\pm = m~ \pa_\tau X^\pm \mp \frac{e}{2} X^\pm
= \mp  \frac{e}{2} x_0^\pm + \frac12 a_0^\pm   e^{\pm e\tau/m}\ ,\quad
\pi^i= m~ \pa_\tau X^i = p^i
\ee
satisfy the usual equal-time commutation rules
\be
\label{compart}
[\pi^+,X^-]=[\pi^-,X^+]=i\ ,\quad [\pi^i,X^j]=i \delta_{ij}
\ee
The world-line Hamiltonian derived from \eqref{lpa}  reads
\be
H = \frac{1}{2m}
\left[ 2 (\pi^+ + \frac{e}{2} X^+)(\pi^- - \frac{e}{2} X^-)
- p_i^2 \right]
= \frac{2a_0^+ a_0^- - p_i^2}{2m}
\ee
and should equal $m/2$ by the mass-shell condition. The 
equation of the trajectory in the light-cone directions may  thus be written
\be
\label{tpa}
(X^+ - x^+_0)(X^- - x^-_0) + \frac{M^2}{2e^2} = 0
\ee
where we denoted by $M^2=m^2 + p_i^2$ the two-dimensional mass.
By convention, we will call particles (or electrons) those following the right
branch of this hyperbola (\ie $a_0^->0$), and 
 anti-particles (or positrons)
those following the left branch (\ie $a_0^-<0$).
The coordinates of the center
of the hyperbola are conserved charges, equal to the
generators of translations in target space
\be
P^\pm =  m \pa_\tau X^\pm \mp e X^\pm = \mp  e x^\pm_0
\ee
The commutation relations \eqref{compart} imply that the two positions
$x^\pm_0$ and velocities $a^\pm_0$ cannot be measured simultaneously,
but rather
\be
\label{partcom}
 [x_0^+, x_0^-] =   -\frac{i}{e} \ ,\quad  [a_0^+, a_0^-] = - i  e
\ee
Finally, one should recall that a constant electric field is
invariant under Lorentz boosts $X^\pm \to e^{\pm \beta} X^\pm$. 
The infinitesimal
generator for this symmetry, commuting with the world-line
Hamiltonian $H$, is
\be
\label{boostpart}
j = X^+ \pi^- - X^- \pi^+ = e x_0^+ x_0^- + \frac{a_0^+ a_0^-}{e}
\ee
where we recall that $a_0^+ a_0^- = M^2/2$. In particular, it
follows from \eqref{tpa} that, just as for neutral particles, 
trajectories of charged particles with $j=0$ go through
the origin $(X^+,X^-)$.

\FIGURE{\label{oneparttraj}
\hfill\epsfig{file=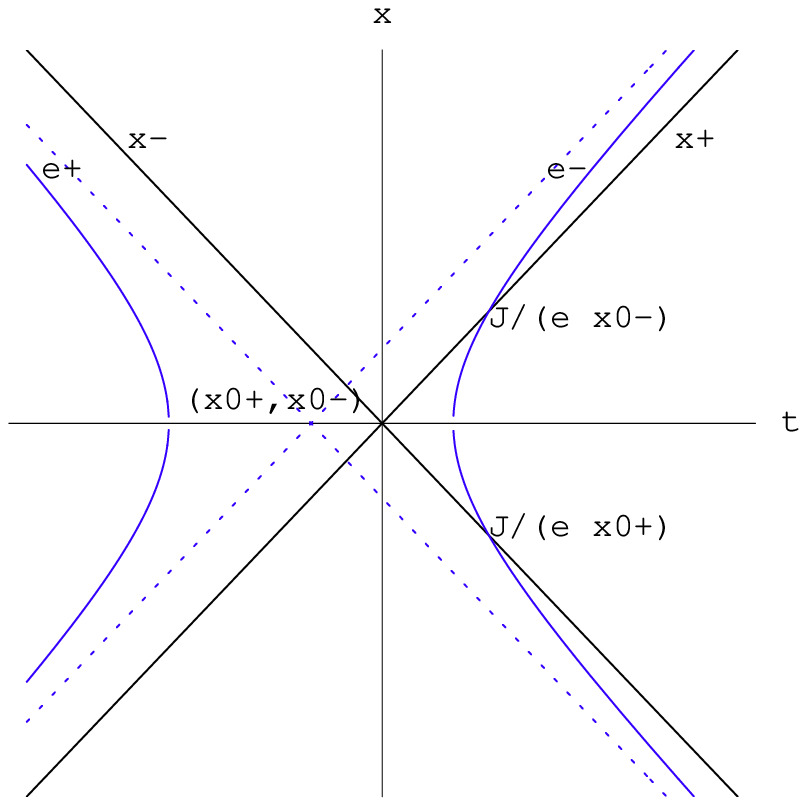,height=9cm}\hfill
\caption{Charged particle in an electric field. The left (right) branch of
the hyperbola represents an positron (electron). The hyperbola is centered
at $(x_0^+,x_0^-)$. The right branch intersects the light cone
at $x^+=j/(e x_0^-)$ and  $x^-=j/(e x_0^+)$. }}

\subsubsection{Vacuum energy and Schwinger pair production}
We now consider the one-loop vacuum free energy associated to a particle
of mass $m$ and spin $s$ in a constant electric field. Using the
Schwinger proper time representation of the propagator, we get
\be
\label{fsch}
{\cal F} = -\frac{1}{2(2\pi)^{D-2}} \int_0^\infty \frac{dt}{t^{D/2}}
\left[ \frac{ e \sinh[(2s+1)e t]}{\sinh^2 e t} -
\frac{2s+1}{t} \right] e^{-i(m^2 - i 0^+)t}
\ee
As is well known,
the free energy density has a non-zero imaginary part, which
can be identified as the production rate of charged pairs \cite{Schwinger:nm}.
The imaginary part can be computed by residues, and reads
\be
\label{schpart}
{\cal W}= \Im {\cal F}
=\frac{1}{2(2\pi)^{D-1}} (2s+1)\sum_{k=1}^{\infty}
(-1)^{F(k+1)} \left(\frac{e}{k}\right)^{D/2}
\exp\left( - \pi k \frac{m^2}{|e|}
\right)
\ee
where the $k$-th term in the sum comes from the pole at
$e t = i k \pi$, and $F=2s \mod 2$ denotes the statistics
of the particle.  As charged particles get created,
they move off to infinity and discharge the condenser plates
which created the electric field. The electric field
therefore relaxes to zero, possibly in a sequence of plasma
oscillations due to recombination phenomena \cite{Cooper:kf,Kluger:1991ib}.

\subsection{Charged string in an electric field}
Let us now recall the basic features of charged open strings propagating
in a constant electric field. We restrict to the bosonic string
for simplicity, and follow \cite{Bachas:bh} except for a few minor
changes in the notations.

\subsubsection{Normal modes and world-sheet Hamiltonian}
The world-sheet action of the bosonic open string in an electric field
reads
\bea
S_{bos} &=& -{1\over 4\pi\alpha'} \int d\sigma  d\tau \  \partial_a X^{\mu}
\partial^a X_{\mu} + \\
&&+ {1\over 2}  \int d\tau \  F_{\mu\nu}^{(0)} X^{\nu}
\partial_{\tau}X^{\mu} \vert_{\sigma=0}\  +\  {1\over 2}  \int d\tau\
F_{\mu\nu}^{(1)}
X^{\nu} \partial_{\tau}X^{\mu} \vert_{\sigma=\pi}
\eea
where $F^{(a)}_{+-}=e_a, a=0,1$ are the values of the electric field at the
two ends of the string, which we may assume to be on different D-branes.
The embedding coordinates are therefore harmonic functions on the world-sheet,
satisfying the boundary conditions
\bea \partial_{\sigma} X^{\pm} &=& \mp \pi e_0 ~\partial_{\tau} X^{\pm}
\quad
(\sigma=0),\\
\partial_{\sigma} X^{\pm} &=& \pm \pi e_1 ~\partial_{\tau}
X^{\pm} \quad
(\sigma=\pi)  \ .
\eea
A simple computation shows that the eigenmodes are integer spaced but
with an imaginary part,
\be
\omega_n=n\pm i\nu\ ,\quad \pi\nu
= \mbox{arcth}(\pi e_0) + \mbox{arcth} (\pi e_1)
\ee
except when the string starts and end on the same brane, whence
$e_0+e_1=0$.
The light-cone embedding coordinates may be expanded in orthonormal modes,
(setting  $\alpha'=1/2$)
\be
\label{omode}
X^{\pm} = x_0^{\pm} + i \sum_{n=-\infty}^{+\infty}
(-)^n (n\pm i\nu)^{-1} a_n^\pm e^{-i(n\pm i \nu) \tau}
\cos[(n\pm i\nu)\sigma \mp i\ \mbox{arcth} (\pi e_0)]
\ee
where our choice of normalization for excited modes
reduces to the standard one \cite{Polchinski:rq} as $\nu\to 0$.
Reality further demands that $(a_n^{\pm})^*= a_{-n}^{\pm}$.
In particular, the zero-modes $a_0^\pm$ and $x_0^\pm$
are hermitian operators. It is worth noting that while each of the points
of the string becomes infinitely accelerated as $\tau\to \infty$,
the length of the string remains finite.

The contribution of the light-cone coordinates $X^\pm$ to the
world-sheet Hamiltonian $L_0$ is now easily computed,
\be
L_0^{l.c.} = - \frac{1}{\pi}
\int_0^\pi \left(\p_\tau X^- \p_\tau X^+ + \p_\sigma X^+ \p_\sigma X^- \right)
= -\frac12 \sum_{m=-\infty}^{\infty} \left(
 a_{-m}^+ a_m^-  + a_{-m}^- a_{m}^+ \right) 
\ee
which we kept Weyl-ordered at this stage, while their contribution
to the higher Virasoro generators reads
\be
L_n^{l.c.} =  - \sum_{m=-\infty}^{\infty}
 a_{-m+n}^+ a_m^-   + L_n^{int}
\ee
We will often assume that the internal conformal field theory is
just flat $\Real^{24}$, or $\Real^8$ in the superstring case.
The canonical commutation relations may be computed easily,
\be [a_m^+, a_n^-] = -(m+i\nu)
\delta_{m+n}\ ,\quad [x_0^+, x_0^-] = - \frac{i}{e_0 + e_1}
\ee
and ensure that $L_0$ generates the time evolution on
the world-sheet, \be [ L_0 , a_{m}^\pm ] = - (m \pm i \nu)
a_{m}^\pm \ee
Note that the operators $x_0^\pm$ and $a_0^\pm$ satisfy
the same algebra as for a charged particle \eqref{partcom},
in the limit of small
electric field, $\nu \sim e_0+e_1$. 
In fact, just as in the particle case $x_0^\pm$ is
proportional to the generators of translations along the
light-cone directions,
\be P^{\pm} = \frac{1}{\pi}\int_0^\pi
\p_\tau X^{\pm} - e_0 X^\pm(0)
-e_1 X^{\pm}(\pi) =  \mp (e_0+e_1)
x^{\pm}_0
\ee
Given these commutation relations, it is natural to try
and quantize the string by assuming the existence of a ground
state annihilated by all strictly positive frequency modes
$a_{n>0}^+, a_{n>0}^-$ and, say, by $a_0^+$. The world-sheet
Hamiltonian, normal ordered with respect to this vacuum,
takes the form
\be
\label{nonrmll}
L_0^{l.c.} = - \sum_{m=0}^{\infty}
a_{-m}^+ a_m^- - \sum_{m=1}^{\infty} a_{-m}^- a_{m}^+ +
\frac{i\nu}{2}(1-i\nu) - \frac{1}{12}
\ee
As we shall see shortly, this quantization prescription is related
by analytic continuation to the prescription for an Euclidean magnetic field.
It is perfectly satisfactory for the purposes of computing
the one-loop vacuum energy, however it does not yield any physical states,
in contrast with classical expectations.
The easily verified hermicitity of $L_0$ does not contradict the fact that
states built on the vacuum by acting with creation operators
$\alpha_{-n}^\pm$ have complex energy, since these states have also
zero-norm, just as in flat Minkowski space.
In the case of the superstring, the fermion vacuum energy in the Ramond sector,
$E_R=1/8-i\nu(1-i\nu)/2$ completely cancels the boson contribution,
while in the Neveu-Schwarz sector, $E_{NS}=-\nu^2/2$
only offsets the quadratic term.

\subsubsection{One-loop amplitude and particle production}
Using this quantization scheme, and freely Wick rotating to
an Euclidean world-sheet, we may now compute the one-loop
vacuum free energy,
\be
\label{onelelec}
A_{bos}=  \frac{i \pi V_{26}
( e_0+ e_1) }{2} \int_0^\infty \frac{dt}{(4\pi^2 t)^{13}}
\frac{e^{-\pi\nu^2 t/2}}{\eta^{21}(it/2)~\theta_1(t\nu/2; it/2)}
\ee
where $\theta_1$ is the Jacobi theta function,
\be
\label{jac}
\theta_1(v;\tau)= 2 q^{1/8} \sin \pi v \prod_{n=1}^\infty
(1-e^{2\pi i v} q^n) ( 1-q^n) (1-e^{-2\pi i v} q^n) \ ,\quad q=e^{2\pi i\tau}
\ee
Just as in \eqref{fsch}, each of the poles  at $t=2k/\nu$ contributes to
the imaginary part, yielding the rate for charged string pair production,
\be
\label{wopen}
{\cal W} = \frac{1}{2(2\pi)^{25}}
\frac{( e_0+e_1)}{\nu} \sum_{k=1}^{\infty} (-)^{k+1}
\left( \frac{ |\nu| }{k} \right)^{13}
\sum_{N=-1}^{\infty} c_b(N) \exp\left(-2\pi k \frac{N}{|\nu|} - 2\pi k |\nu| 
\right)
\ee
where we expanded $\eta^{-24}(q)=\sum_{N=-1}^{\infty} c_b(N) q^N$.
The result \eqref{wopen} 
can be viewed as the sum of the production rates \eqref{schpart} for
each state in the string spectrum, of mass $m^2 = 2N + \nu^2$.
Note that the poles originate from the $\sin(\pi \nu t /2)$ term
in \eqref{jac}, which represents the contribution of the zero-modes.

Similarly, the one-loop amplitude in the oriented superstring case reads
\be A_{ferm}= \frac{i \pi V_{10}
(e_0+e_1)}{2} \int_0^\infty \frac{dt}{(4\pi^2 t)^{5}}
\frac{\theta_1^4(t\nu/4; it/2)}{\eta^{9}(it/2)~\theta_1(t\nu/2;it/2)}
\ee
The imaginary part can be computed as above, and yields
\be {\cal W} = \frac{1}{2(2\pi)^{9}}
\frac{e_0+e_1}{\nu} \sum_{k=1, k ~\mbox{\scriptsize odd}}^{\infty}
\left( \frac{
|\nu| }{k} \right)^{5}
\sum_{N=0}^{\infty} c_f(N) e^{-2\pi k N/|\nu|}
\ee
where $(\theta_3^4-\theta_4^4)/\eta^{12}(q)=\theta_2^4(q)/\eta^{12}(q)=
\sum_{N=0}^{\infty} c_f(N) q^N$ is the partition function in the
Neveu-Schwarz sector, equal to that in the Ramond sector.
This is in agreement with Schwinger's formula \eqref{schpart},
upon identifying $m^2=2N$. In particular, there is no quadratic mass shift,
despite the occurrence of such a term in the vacuum energy of $L_0$
in the Neveu-Schwarz sector: the contribution of the
massless sector of the open superstring therefore is only
power suppressed as a function of $k$,  and converges only when
sufficiently many non-compact dimensions are included. Note also
that the contributions with even Schwinger index $k$ cancel
between bosons and fermions, as a consequence of supersymmetry\footnote{In
this sense, introducing an electric field breaks SUSY only softly.}.

\subsubsection{Normal ordering and physical states}
Let us now return to our normal ordered expression
\eqref{nonrmll} for $L_0$. While it leads to a one-loop amplitude
\eqref{onelelec}
with a satisfactory interpretation,
it suffers from a major drawback: the energy of any state
constructed by applying creation operators
on this vacuum has an imaginary part which is an odd
multiple of $i\nu/2$, and therefore never satisfies the
physical condition $L_0=0$ \cite{Nekrasov:2002kf}.
One may object that a uniform
constant electric field is not physical anyway, since
infinite energy may be transferred to charged particles;
furthermore, a realistic electric field could only occur in a finite region 
of space bounded by condenser plates or other device that creates it. 
This answer is not satisfactory, as one may well scatter electrons
and positrons in a finite region immersed in an electric field,
of size much larger than the wavelength of the incoming particles.
Clearly there is no problem in setting this up classically,
and a quantum induced vacuum energy should not cause
electrons to vanish into inconsistency.

Let us analyze the reason for this imaginary vacuum energy.
The Fock space condition $a_0^+ | 0 \rangle = 0$ is simply the
analytic continuation of the $a^z_0 | 0 \rangle = 0$ condition
which picks out the normalizable eigenstates of the two-dimensional harmonic
oscillator describing the Landau orbits of charged particles
in a constant magnetic field: upon analytically continuing $B \to i \nu$
so as to obtain an electric field,
these states become eigenstates of an {\it inverted} harmonic
oscillator with imaginary
energy $i(n+1/2)\nu$.
Nevertheless, the inverted harmonic oscillator admits
a perfectly well behaved (although unbounded) spectrum
of delta-normalizable scattering states
with real energy: those are the ones which describe the unbounded trajectories
of the electron and positron in an electric field. Upon continuation
back to the Euclidean magnetic case, these states would correspond
to negative energy non-normalizable states.

Given that the inverted harmonic oscillator does not have a ground state,
there is no reason to normal order in the zero-mode sector, and
we therefore write
\be \label{nrmll} L_0^{l.c.} = -
\frac12 \left( a_0^+ a_0^- + a_0^- a_0^+ \right) -
\sum_{m=1}^{\infty} \left( a_{-m}^+ a_m^-  + a_{-m}^- a_{m}^+
\right) + \frac12 \nu^2  - \frac{1}{12} \ee
This generator now has a real vacuum energy in the excited sectors.
We follow the same ordering rule for the generator of boosts,
\bea
\label{boostopen}
J &=&  \frac{1}{\pi}\int_0^\pi \left(X^+ \p_\tau X^- - X^- \p_\tau X^+ \right)
+ \frac{e_0}{\pi} X^+ X^-(0) + \frac{e_1}{\pi} X^+ X^-(\pi) \\
&=&\frac12 (e_0+e_1) \left(x_0^+ x_0^- + x_0^- x_0^+\right)
+\frac{1}{2\nu} \left(  a_{0}^+ a_0^- + a_0^- a_{0}^+ \right)
+i \sum_{m=1}^{\infty}  \left(
\frac{ a_{-m}^- a_m^+}{m+i\nu} -
\frac{ a_{-m}^+ a_{m}^-}{m-i\nu} \right)\nn 
\eea
where the zero-mode contribution reproduces the boost momentum $j$
of a charged particle \eqref{boostpart}.
We will discuss
in Section 3 the physical content of this ordering prescription.


\subsection{Twisted closed strings in the Milne Universe}
Let us now turn to the case of closed strings propagating in
flat two-dimensional Minkowski space orbifolded by a discrete
Lorentz boost $X^\pm \equiv e^{\pm\beta} X^\pm$. We recall that this
space can be interpreted as a cosmological universe with
a circle contracting linearly with time until a Big Crunch singularity,
followed by a linear Big Bang expansion:
\be
ds^2 = - 2 dX^+ dX^- = -dT^2 +  T^2 d\theta^2
\ee
where the Milne coordinates $X^\pm = T e^{\pm \theta}/\sqrt{2}$
cover the forward ($T>0$) and past ($T<0$) light-cone of two-dimensional
Minkowski space. The effect of the orbifold is to make the
spatial coordinate compact, $\theta\equiv \theta+\beta$.
In addition, the left and right quadrants of the covering space
descend two regions attached to the singularity at $T=0$, with metric
\be
 ds^2 = - 2 dX^+ dX^- = - r^2 d\eta^2 + dr^2
\ee
where $X^\pm = \pm r e^{\pm \eta}/\sqrt{2}$ parameterize
Rindler space. The orbifold action identifies the Rindler time
under\footnote{Upon continuing $\beta$ to $i\beta$, one thus obtains the
idenfitication appropriate for thermal Rindler space.}  
$\eta\equiv \eta+\beta$ hence generates CTC's.
It will be also useful to define $r=\pm e^y, T=\pm e^{\tau}$
(with appropriate signs in each patch) so that
the metric becomes conformal to a cylinder,
\be
ds^2=e^{2y}(-d\eta^2 + dy^2) = e^{2\tau} (-d\tau^2 + d\theta^2)
\ee
with the singularity being pushed at $y=-\infty$ or $\tau=-\infty$.
Finally, the light-cone $X^+X^-=0$ on the covering space descends
to a non-Hausdorff null space lying arbitrarily close
to the cosmological singularity.

While the Lorentzian orbifold exhibits a cosmological singularity and
CTC's, it has been suggested to resolve these problems
by combining the boost with a translation
on an extra spatial direction \cite{Cornalba:2002fi,
Cornalba:2002nv,Cornalba:2003ze}. It is unclear however whether this
deformation survives quantum dynamics, or relaxes dynamically
to the singular configuration.
Upon reduction along the direction where the translation is performed,
this ``electric Melvin universe'' can be interpreted
as an electric background for the Kaluza-Klein gauge field, on a
curved geometry induced by the electric energy. Closed time-like curves
still exist but are shielded behind a Cauchy horizon.
Kaluza-Klein modes along the compact circle are charged under the electric
field, and so may be pair-produced \cite{Friedmann:2002gx}. Our
concern in this paper however is in the kinematics of twisted strings,
which are neutral under this gauge field, and insensitive to the
deformation.

\subsubsection{Normal mode expansion}
As in any orbifold, closed strings fall into different twisted sectors
depending how many times they wind around the space-like circle $S^1_\theta$
in the past and forward patches, or the time-like circle $S^1_\eta$ in the
left and right patches. Strings in the untwisted sector have the
standard free mode expansion, but for the condition that they should
have integer total momentum along $\theta$, or energy along $\eta$.
Closed strings in the $w$-th twisted sector on the other hand satisfy the
twisted periodicity condition
\be
X^{\pm}(\sigma+2\pi,\tau)=e^{\pm w \beta} X^{\pm}(\sigma,\tau)
\ee
and hence have complex proper frequencies $\omega_n=n\pm i\nu$.
This is analogous to the spectrum of open strings upon identifying the 
product of the boost parameter by the winding number with the electric
field, according to
\be
\label{idee}
w \beta = - 2\pi \nu= -2 \mbox{arcth}(\pi e_0)\ ,\qquad e_1=0\ .
\ee
The string embedding coordinates may be
expanded in left and right moving eigenmodes
\be
X^\pm_{closed}=X_R^\pm(\tau-\sigma) + X_L^\pm(\tau+\sigma)
\ee
where, in analogy with \eqref{omode},
\bea
\label{closedmod}
X_R^\pm(\tau-\sigma) &=& \frac{i}{2} \sum_{n=-\infty}^{\infty}
(n \pm i\nu)^{-1} \alpha_n^\pm e^{-i(n\pm i \nu)(\tau-\sigma)} \\
X_L^\pm(\tau+\sigma) &=& \frac{i}{2} \sum_{n=-\infty}^{\infty}
(n \mp i\nu)^{-1}
\tilde \alpha_n^\pm e^{-i(n\mp i \nu)(\tau+\sigma)}
\eea
The closed string oscillators satisfy the following
commutation relations and 
hermiticity properties
\bea
[\alpha_m^+, \alpha_n^-] =
-(m+i\nu) \delta_{m+n}\ &,&\quad
[\tilde \alpha_m^+, \tilde \alpha_n^-] =
-(m-i\nu) \delta_{m+n}\ ,\quad
\\
(\alpha_{-n}^\pm)^* = \alpha_n^\pm\ &,&\quad (\tilde
\alpha_{-n}^\pm)^* = \tilde \alpha_n^\pm \eea
The left and right-moving
world-sheet Hamiltonians are easily found to be
\bea
\label{closedls} L_0^{l.c} &=& -\frac12 ( \alpha_0^+ \alpha_0^- +
\alpha_0^- \alpha_0^+ ) -\sum_{n=1}^{\infty}
\left( \alpha_{-n}^+ \alpha_n^- +  \alpha_{-n}^-\alpha_n^+ \right)
+ \frac12 \nu^2 -\frac{1}{12}\\
\tilde L_0^{l.c.} &=&  -\frac12 ( \tilde \alpha_0^+ \tilde \alpha_0^-
+ \tilde \alpha_0^- \tilde \alpha_0^+ )
-\sum_{n=1}^{\infty}  \left( \tilde \alpha_{-n}^+\tilde \alpha_n^-
+ \tilde \alpha_{-n}^- \tilde \alpha_n^+ \right)
+ \frac12 \nu^2 -\frac{1}{12}
\eea
where we followed the same normal ordering prescription as in the open string
case. Again, this prescription is needed to obtain non trivial physical
states, and will be discussed at length in Section 3.
We follow the same rule in displaying the generator of boosts,
\bea \label{boostclosed} J=
\frac{1}{\pi}\int_0^{2\pi} && \left(X^+ \p_\tau X^- - X^- \p_\tau
X^+ \right) = \frac{1}{2\nu} \left(
 \alpha^+_0\alpha^-_0 +
\alpha^+_0\alpha^-_0 \right)
-\frac{1}{2\nu} \left( \talpha^+_0\talpha^-_0 + \talpha^-_0\talpha^+_0
\right) \nn \\
&& +~i \sum_{m=1}^{\infty} \left(
\frac{\alpha_{-m}^-  \alpha_{m}^+ - \talpha_{-m}^+  \talpha_{m}^- }{m+i\nu}
+\frac{\talpha_{-m}^-  \talpha_{m}^+  -\alpha_{-m}^+  \alpha_{m}^- }{m-i\nu}
 \right)
\eea
Physical states satisfy $L_0=\tilde
L_0=0$. In contrast to the open string case, the orbifold projection
demands that the total boost momentum $J$ be an integer multiple of 
$2\pi/\beta$.
In particular, transverse massless states necessarily have $J=0$,
which as we shall see in Section \ref{semirate} localizes particle
production to the light-cone.

\subsubsection{Open vs closed strings}
As a matter of fact, the open string mode expansion \eqref{omode}
may be obtained from \eqref{closedmod} by identifying the
left and right-moving oscillators as
\be
\alpha_n^\pm= (-)^n a_n^\pm e^{\pm \mbox{arcth}(\pi e_0)}
= - \tilde\alpha_{-n}^\pm
\ee
and introducing an extra zero-mode $x_0^\pm$, so that
\be
X_{open}^{\pm} = x_0^\pm + 
X_R^\pm(\tau-\sigma) + 
X_L^\pm(-\tau-\sigma)\ ,\quad
\ee
Despite the fact that the twisted closed string has twice as many excited modes
as the charged open string, it is important to note that the zero-mode
structures are in fact identical.  The closed string possesses two pairs
of level-zero oscillators $\alpha_0^\pm$ and $\tilde\alpha_0^\pm$, while
the open string has two pairs of oscillators $a_0^\pm$ and $x_0^\pm$.
The two satisfy the same algebra \eqref{partcom} upon identifying
\be
\label{oczero}
\alpha_0^\pm = a_0^\pm\ ,\qquad \talpha_0^\pm =
\pm \sqrt{\nu(e_0+e_1)} x_0^\pm
\ee
They differ in the way they appear in the world-sheet Hamiltonian
$L_0+\tilde L_0$ however. Vertex operators at the zero-mode level
can nevertheless be treated identically in the open and closed string case.



\subsubsection{One-loop amplitude\label{cclose}}
The one-loop amplitude for the Milne orbifold has been computed
in \cite{Nekrasov:2002kf,Cornalba:2002fi}, using an Euclidean
world-sheet but a Lorentzian target space. In the bosonic string
case, the result can be written as an integral over the fundamental
domain of the modular group
\be
\label{milne1l}
A_{bos}=\int_{{\cal F}}
\sum_{l,w=0}^{\infty} \frac{d\tau d\bar\tau}{(2\pi^2 \tau_2)^{13}}
\frac{e^{-2\pi  \beta^2 w^2 \tau_2}}
{\left| \eta^{21}(\tau)
~\theta_1(i \beta(l+w\tau); \tau) \right| ^2 }
\ee
whereas for the fermionic string,
\be
\label{milne1s}
 A_{ferm}=\int_{{\cal F}}
\sum_{l,w=0}^{\infty} \frac{d\tau d\bar\tau}{(2\pi^2 \tau_2)^{5}}
\left| \frac{\theta_1^4(i \beta(l+w\tau)/2; \tau)}{
\eta^9(\tau) ~\theta_1(i \beta(l+w\tau); \tau)}\right| ^2
\ee
In either expression, the
integers $l,w$ denote the winding numbers along the two
cycles of the torus. The partition functions \eqref{milne1l}
and \eqref{milne1s}
are indeed modular invariant,
and agree with the standard quantization prescription
based on a Fock vacuum annihilated by the oscillators
$\alpha^{\pm}_{m>0}, \talpha^{\pm}_{m>0}, \alpha_0^+$ and $\talpha_0^+$.
In contrast to open string case however, the physical interpretation
of the integrated amplitude is obscured by the existence of poles in
the bulk of the fundamental domain. Indeed, the result \eqref{milne1l}
is very similar \cite{Cornalba:2002fi} to 
the one-loop amplitude for the Euclidean BTZ black
hole \cite{Maldacena:2000kv}. In the BTZ case, the divergences 
were interpreted as contributions from continuous representations 
of affine $Sl(2)_k$, corresponding
to long strings extending to the boundary of $AdS_3$.
Whether long strings play a r\^ole in the Milne orbifold as well
will be left to future  work.

\section{First quantization, revisited}
As we have seen, both the charged open string and the twisted
closed string in the Milne Universe behave like a charged massive
particle in a constant electric field, from the point of view of
their zero-mode degrees of freedom. In Section 3.1, we represent
the zero-modes on the space of wave functions of the center of
motion. In Section 3.2 we study in
detail the wave functions of the charged Klein-Gordon equation,
which provide the first approximation to the string vertex
operators. We illustrate in Section 3.3 our prescription
on the spectrum of low-lying physical states in the framework 
of old covariant quantization, and reconcile in Section 3.4 our
quantization prescription with the one-loop amplitude.

\subsection{Quantizing the zero-modes}

\subsubsection{Open strings in constant electric field}

As we have seen, the charged open strings and twisted closed
strings have isomorphic zero-modes structures. 
It is therefore important to have a
complete physical understanding of these zero-modes. Let us
therefore go back to the open string case
\be \label{os2}
[x_0^+,x_0^-] = - \frac{i}{e_0+e_1}\ ,
\quad [a_0^+,a_0^-] = -i \nu
\ee
with the two pairs of oscillators commuting.
The relation of these operators to more familiar position and
momenta may be obtained by looking at the limit of small
electric field $\nu \sim (e_0+e_1) \to 0$. The mode expansion
is singular in this limit, 
\be X^\pm \to \left(x_0^\pm \pm
\frac{a_0^\pm}{\nu} \right) + a_0^\pm \tau ~+~ \mbox{osc.} 
\ee
One should therefore identify the position and velocity operators as
\be x^{\pm} = x_0^\pm \pm
\frac{a^\pm}{\nu} \ ,\quad p^\pm = a_0^\pm \ee
 The commutation
relations \eqref{os2} guarantee that they satisfy the correct
commutation relation $[x^\mu,p^\nu]=i\eta^{\mu\nu}$ in the limit
(where $\eta^{+-}=-1$). The velocity operators $a_0^\pm=p^{\pm}$
may therefore be represented in the space of functions of the
target space coordinates $(x^+,x^-)$ as $p^\pm \sim i \p/\p x^\mp$, up
to corrections that vanish as $\nu\to 0$. It turns out that $a_0^\pm$
can be identified with the covariant derivatives
in a constant electric field,
\be
\label{cov}
a_0^\pm = p^\pm =
i \p_\mp \pm \frac{\nu}{2} x^\pm \ ,\quad x_0^\pm = \mp
\frac{1}{\sqrt{\nu( e_0 + e_1)}} \left( i \partial_{\mp}
\mp \frac{\nu}{2} x^\pm \right)
\ee
 These operators are hermitian
with respect to the $L_2$ norm on target space $\int d\bar x^+d\bar x^-
|f|^2$. The mass-shell condition $L_0=0$ now becomes the Klein-Gordon
equation for a charged particle in a constant  electric field
$\nu$,
\be 
\label{kg}
L_0 = -\frac12 \left[ \left( i\p_+ + \frac{\nu}{2}
x^- \right) \left( i\p_- - \frac{\nu}{2} x^+ \right) + \left(
i\p_- - \frac{\nu}{2} x^+ \right) \left( i\p_+ + \frac{\nu}{2}
x^- \right) \right] + \frac12 M^2 \equiv 0
\ee
 where
 \be M^2:=a_0^+ a_0^- +
a_0^- a_0^+ = - 2 \sum_{m=1}^{\infty} \left( a_{-m}^+ a_m^-  +
a_{-m}^- a_{m}^+ \right) + \nu^2  - \frac{1}{6} + 2  L_0^{int}
\ee
denotes the two-dimensional mass squared. The zero-modes
($a_0^\pm,x_0^\pm)$
therefore simply represent the degrees of freedom of the center of
mass of the charged string. In this representation, the generator
of boosts takes the simple form
\be J= -\frac{i}{2} \left(x^+
\p_+ -x_- \p_- \right) +i
\sum_{m=1}^{\infty}
\left(\frac{  a_{-m}^- a_{m}^+ }{m+i\nu}
-\frac{a_{-m}^+ a_m^-}{m-i\nu} \right)
\ee

\subsubsection{Closed strings in the Milne Universe}
Similarly, the  closed string zero-modes are described by a pair of
hermitian canonical conjugate variables, satisfying
\be
\label{czero}
\left[ \a_0^+, \a_0^- \right] = -i \nu\ ,\quad
\left[ \ta_0^+, \ta_0^- \right] = i \nu\ .
\ee
Their meaning may be understood in the flat space limit
\be \nu\rightarrow 0,\ \ \ x^+\rightarrow
{-R\over\nu}+y^+,\ \ \ x^-\rightarrow
{-R\over\nu}+y^-\ ,\quad y^\pm \ \mbox{finite}
\ee
where the Milne geometry reduces to a finite circle of
constant radius $R$, times the time direction.
The string embedding coordinates reduce to
\be X^\pm\rightarrow \pm {\a_0^\pm +\ta_0^\pm \over \nu }
\pm (\a_0^\pm -\ta_0^\pm)\tau -
(\a_0^\pm+ \ta_0^\pm)\sigma + \mbox{osc.}
\ee
so that  $\a_0^\pm\ , \ta_0^\pm$ are related to the closed string zero-modes
in flat space $y_0^\pm, p^\pm$ by the singular field redefinition,
\be
y_0^\pm = \frac{1}{\nu}(\a_0^\pm + \ta_0^\pm \pm R)\ ,\quad
p^\pm=\pm (\a_0^\pm - \ta_0^\pm)
\ee
One may easily verify that the
commutation relations \eqref{czero}
contract to the usual relations for the zero-modes on $\Real\times S^1$,
in the limit $\nu\to 0$ with fixed $R$.

Thanks to the isomorphism \eqref{oczero}, or on the basis of
this identification, the same representation \eqref{cov}
in terms of differential operators acting on the space of wave functions
$\phi(x^+,x^-)$ can be used to
describe the zero-mode degrees of freedom of the twisted closed string:
\be
\alpha_0^\pm =  i \partial_{\mp} \pm \frac{\nu}{2} x^\pm\ ,\quad
\talpha_0^\pm = i \p_\mp \mp \frac{\nu}{2} x^\pm \ee
The zero-mode piece of $L_0$ and $\tilde L_0$
\be {\cal
M}^2:=\alpha_0^+ \alpha_0^- +  \alpha_0^- \alpha_0^+\ ,\quad
\tilde {\cal M}^2:=\talpha_0^+ \talpha_0^- +  \talpha_0^-
\talpha_0^+ \ee
are therefore the Klein-Gordon equation of a
particle of mass squared and charge $({\cal M}^2,\nu)$ and
$(\tilde {\cal M}^2,-\nu$), respectively, where
\bea {\cal M}^2
&=&  -  \sum_{n=1}^{\infty} \left( \alpha_{-n}^+ \alpha_n^- +
\alpha_{-n}^-\alpha_n^+
\right) + \frac12\nu^2 -\frac1{12} + L_{int} \\
\tilde {\cal M}^2 &=&  -  \sum_{n=1}^{\infty} \left(
\talpha_{-n}^+ \talpha_n^- +  \talpha_{-n}^-\talpha_n^+ \right) +
 \frac12\nu^2 -\frac1{12} +\tilde L_{int}\eea
 Their difference is
equal to the zero-mode Rindler boost momentum $j$,
\be {\cal M}^2 -
\tilde {\cal M}^2 = 2i\nu\left( x^+ \p_+ - x^- \p_- \right)
=2\nu j
\ee
The level matching condition $L_0-\tilde L_0=0$ relates the
product of the boost momentum $J$ by the winding $w$ to the
difference of occupation numbers of the excited levels,
\be 
\label{matchm}
\tilde L_0 - L_0
= \nu J +\sum_{m=1}^{\infty} \frac{m}{m+i\nu}
\left(\alpha_{-m}^- \alpha_m^+ - \talpha_{-m}^+ \talpha_{m}^- \right)
+\sum_{m=1}^{\infty} \frac{m}{m-i\nu}
\left(\alpha_{-m}^+ \alpha_m^- - \talpha_{-m}^- \talpha_{m}^+ \right)
\ee
Notice that the right-hand side is integer valued as it
should. The orbifold projection on the other hand requires that
the total boost momentum $J$ be integer,
\be \label{boostclosed2}
J=
-i (x^+ \p_+ -x^- \p_-) +~i \sum_{m=1}^{\infty} \left(
\frac{\alpha_{-m}^-  \alpha_{m}^+ - \talpha_{-m}^+  \talpha_{m}^- }{m+i\nu}
+\frac{\talpha_{-m}^-  \talpha_{m}^+  -\alpha_{-m}^+  \alpha_{m}^- }{m-i\nu}
 \right)
\in \frac{2\pi}{\beta} \Zint \ee
We have therefore obtained our main insight into the
dynamics of twisted closed strings in the Milne universe: from the
point of view of their center of mass, they behave like particles
of charge $w$ in a constant electric field $e$ related to the
boost parameter $\beta$ by Eq. \eqref{idee}, 
with a further restriction on the boost momentum $J$ as specified
by the matching relation.

\subsection{Zero-mode wave functions in static coordinates}
In this section, we quantize the charged Klein-Gordon equation \eqref{kg}
governing the zero-mode degree of freedom from the point
of view of a static observer, who measures scattering processes with respect
to the global time $X^0=t$.

\subsubsection{The inverted harmonic oscillator}
In order to go to static coordinates, we define linear combinations
\be
a^\pm_0=\frac{P\pm Q}{\sqrt{2}}, \quad
x^{\pm}_0=
\frac{\tilde P\pm \tilde Q}{\sqrt{2\nu(e_0+e_1)}}
\ee
satisfying the canonical commutation relations
\be
[P,Q]=i\nu\ ,\quad [\tilde P,\tilde Q]=-i\nu
\ee
Equivalently, using the representation of $a_0^\pm,x_0^\pm$
as covariant derivatives,
\be
P=i\p_t + \frac{\nu}{2} x\ ,\quad
Q=-i\p_x + \frac{\nu}{2} t\ ,\quad
\tilde P=i\p_t - \frac{\nu}{2} x\ ,\quad
\tilde Q=-i\p_x - \frac{\nu}{2} t
\ee
The charged Klein-Gordon operator can now be rewritten as an inverted harmonic
oscillator
\be
- M^2 = P^2 - Q^2
\ee
Since the generator of spatial translations $\tilde P$ commutes with $M^2$,
we may diagonalize it and further compute the wave functions
in the $P$ representation. Since $P=\tilde P+\nu x$, we expand
\be
f(x^+,x^-)=\int d\tilde p ~ \psi_{\tilde p}(u)
e^{-i (\tilde p + \frac12 \nu x) t }
\ee
where we defined $u=(\tilde p+\nu x)\sqrt{2/\nu}$. The Klein-Gordon
equation now takes the form of an inverted harmonic oscillator in one
variable,
\be
\label{invu}
\left(-\pa_u^2 - \frac14{u^2} + \frac{M^2}{2\nu}
\right) \psi_{\tilde p}(u) = 0
\ee
The physical interpretation of the motion in the inverted harmonic
potential is now clear: particles coming from
$u=+\infty$ in the inverted harmonic potential are just electrons
coming from $x=+\infty$ and being slowed down by the electric field.
At the turning point $u=M\sqrt{2/\nu}$ or $x=(M-\tilde p)/\nu$,
they bounce against the potentiel barrier and escape to $x=+\infty$
again. In real space, the trajectory consists of a branch of an
hyperbola centered at $x=-\tilde p/\nu$. Similarly, particles
coming from $u=-\infty$ are positrons which follow the other
branch of the hyperbola.

Quantum mechanically, there therefore exists a continuum of
delta-function normalizable states with real positive $M^2$,
in stark contrast with expectations based on analytic
continuation from the Euclidean magnetic problem. Furthermore,
the quantum mechanical tunnelling through the barrier can be simply
interpreted as induced Schwinger pair production
$e^- \to (1+\eta) e^- + \eta e^+$. We will compute the 
production rate $\eta$ in the next subsection. Notice that the
fact that the potential is unbounded from below does not
cause any instability, as the energy is fixed equal to $M^2/(2\nu)$.
On the other hand, the Schwinger pair production will be responsible
for a dynamical decay of the electric field, as the charged particles
move off to infinity and screen the electric field.


\subsubsection{Parabolic cylinder functions}

Eigenmodes of the inverted harmonic oscillator are well known
under the name of parabolic cylinder functions. They already
arose in the context of the $c=1$ string theory  \cite{Moore:1991sf}, 
from where
we borrow the following results: even
and odd solutions of \eqref{invu} under the parity transformation
$u\to -u$ are given by
\bea
\psi^+(u) &=&
\frac{2^{1/4}}{\sqrt{4\pi(1+e^{\pi M^2/\nu})^{1/2}}}
\left|
\frac{\Gamma(\frac14+i \frac{M^2}{4\nu})}
{\Gamma(\frac34+i \frac{M^2}{4\nu})}
\right|^{1/2}
e^{-i u^2/4} ~{}_1 F_1\left(\frac14-i \frac{M^2}{4\nu},
\frac12; i \frac{u^2}{2}\right) \nn\\
\psi^-(u) &=& \frac{2^{3/4}}{\sqrt{4\pi(1+e^{\pi M^2/\nu})^{1/2}}}
\left|
\frac{\Gamma(\frac34+i \frac{M^2}{4\nu})}{\Gamma(\frac14+i \frac{M^2}{4\nu})}
\right|^{1/2}
u~ e^{-i u^2/4} ~{}_1 F_1\left(\frac34-i \frac{M^2}{4\nu}, \frac32
; i \frac{u^2}{2}\right) \nn
\eea
These form a delta-normalizable complete orthogonal basis of functions
on $\Real$, normalized such that
\bea
\sum_{\eps=\pm} \int_{-\infty}^{\infty}
\frac{dM^2}{2\nu} ~\psi^\eps(M^2,u_1)\psi^\eps(M^2,u_2) &=& \delta(u_1-u_2) \\
\sum_{\eps=\pm} \int_{-\infty}^{\infty}
du ~\psi^\eps(M_1^2,u)\psi^\eps(M_2^2,u) &=& 2 \nu\delta(M_1^2-M_2^2)
\eea
For large positive $u\gg M/\sqrt{\eps}$, they satisfy the BKW asymptotics
\be
\label{bkw}
\psi^\pm(u)
\sim {\left( 2\pi u\sqrt{1+e^{\pi M^2/\nu}} \right)^{-1/2}}
\left[ \sqrt{k} \cos \phi(u)
\pm \frac{1}{\sqrt{k}} \sin\phi(u) \right]
\ ,
\ee
where
\be
\phi(u) = \frac14 u^2 -\frac{M^2}{\nu} \log |u|
+\frac{\pi}{4}+
\frac12 \arg \Gamma\left(\frac12 + i \frac{M^2}{2\nu} \right)
\ee
and $k=\sqrt{1+e^{\pi M^2/\nu}}-e^{\pi M^2/\nu}$.
In particular they fall off at infinity as $1/\sqrt{|u|}$.

\subsubsection{In and out states}
It is now possible to construct plane-wave combinations,
such that only incoming or outgoing waves are present at $u=\pm \infty$:
\bea
\psi_R^\pm(u) &=& \frac12 (k^{-1/2}\pm i k^{1/2}) \psi^+
+ \frac12 (k^{-1/2}\mp i k^{1/2}) \psi^- \\
\psi_L^\pm(u) &=& \frac12 (k^{-1/2}\pm i k^{1/2}) \psi^+
- \frac12 (k^{-1/2}\mp i k^{1/2}) \psi^-
\eea
such that
\bea
\psi_R^\pm(u) &\sim& {\left( 2\pi u\sqrt{1+e^{\pi M^2/\nu}} \right)^{1/2}}
e^{\pm i \phi(u) } \ ,\quad u \to +\infty\\
\psi_L^\pm(u) &\sim&  {\left( -2\pi u\sqrt{1+e^{\pi M^2/\nu}} \right)^{1/2}}
e^{\pm i \phi(u)} \ ,\quad u \to -\infty
\eea
These combinations can in fact be written directly
in terms of parabolic cylinder functions
$D_{-\frac12\pm i\frac{M^2}{2\nu}}(\pm e^{\pi 3i\pi/4}u)$
(see Appendix \ref{cylpar}).
Using these modes we may now construct
solutions of the Klein-Gordon equation
corresponding to the creation of an electron or the
annihilation of a positron at $t=-\infty$,
\bea
\phi^{in}_e = \psi_R^- (u) e^{-i(\tilde p + \frac12 \nu x)t}\ ,\quad
\phi^{in}_p = \psi_L^+ (u) e^{-i(\tilde p + \frac12 \nu x)t}
\eea
or at $t=+\infty$:
\bea
\phi^{out}_e = \psi_R^+ (u) e^{-i(\tilde p + \frac12 \nu x)t}\ ,\quad
\phi^{out}_p = \psi_L^- (u) e^{-i(\tilde p + \frac12 \nu x)t}
\eea
The semi-classical behaviour of these wave functions is summarized
on Figure \ref{staticfig}. The reflection and transmission coefficients
of the wave function 
amplitude are easily computed by looking at the semi-classical
expansion in the region $u\to -\infty$,
\be
\begin{pmatrix}
\psi^+_R\\
\psi^-_R
\end{pmatrix}(u)
 \sim  \frac{i}{2} {\left( -2\pi u\sqrt{1+e^{\pi M^2/\nu}} \right)^{1/2}}
\begin{pmatrix}
k-\frac{1}{k} & k+\frac{1}{k}  \\
-k-\frac{1}{k} & -k+\frac{1}{k}
\end{pmatrix}
\begin{pmatrix}
e^{i \phi(u)}\\
e^{-i \phi(u)}
\end{pmatrix}
\ee
which yield
\be
R= \frac{k^{-1}+k}{k^{-1}-k} =
-\frac{1}{\sqrt{1+e^{-\pi M^2/\nu}}}\ ,\quad
T= \frac{2}{k-k^{-1}} = \frac{1}{\sqrt{1+e^{\pi M^2/\nu}}}\ ,\quad
\ee
The fact that the modulus of the reflection coefficient is greater
than one is a manifestation of the Klein paradox, due to the unboundedness
of the potential. As usual, the resolution of the paradox is that the charge 
density $q=u|\psi(u)|^2$ is proportional 
to the modulus square of the
amplitude, and conserved during the stimulated pair production process
$e^- \to R^2 e^- + T^2 e^+$ 
thanks to the relation $R^2 = 1 + T^2$. The stimulated 
pair creation rate $\eta$ introduced at the end of Section 3.2.1
is therefore equal to $\eta=R^2-1$.

\FIGURE{ \label{staticfig}
\hfill\epsfig{file=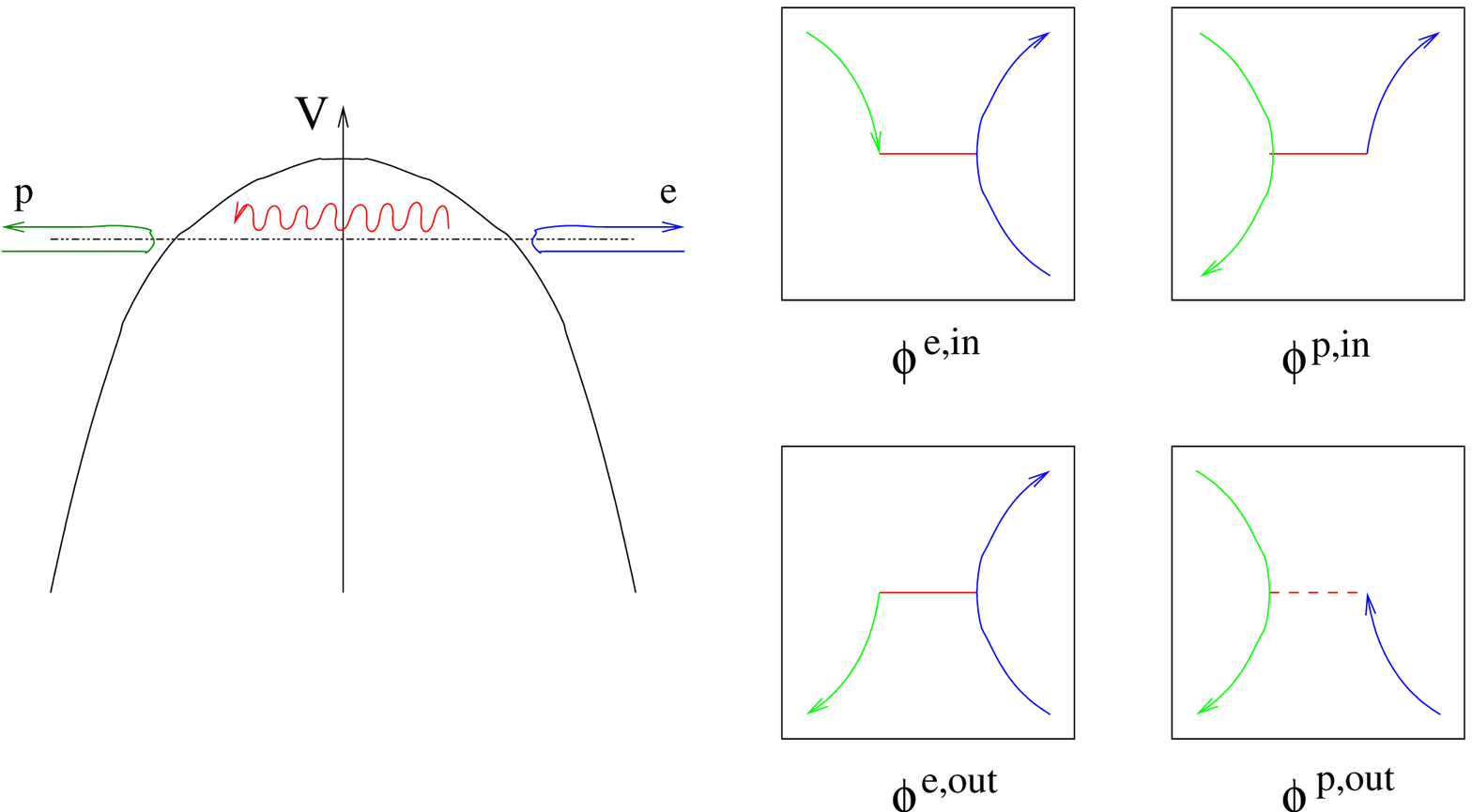,height=8cm}
\hfill
\caption{Creation and annihilation modes in static coordinates.}}

A general solution of the Klein-Gordon equation
may therefore be expanded in two different ways,
\be
\phi = \int d\tilde p ~\left(
a_{\tilde p}^{in*} ~\phi^{\tilde p,in}_e
+b_{\tilde p} ~\phi^{\tilde p,in}_p \right)
=\int d\tilde p ~\left(
a_{\tilde p}^{out*} ~\phi^{\tilde p,out}_e
+b_{\tilde p}^{out} ~\phi^{\tilde p,out}_p \right)
\ee
Here the operator $a^*$ creates an electron, whereas $b^*$ creates a
positron. The commutation relations read, either in the
{\it in} or {\it out} basis,
\be
[a_{\tilde p}, a^{*}_{\tilde p'}] = [b_{\tilde p}, b^{*}_{\tilde p'}]
=\delta(\tilde p-\tilde p')
\ee
The non-trivial Bogolioubov transformation between
these two bases is a consequence of pair production in an electric
field. Using the identities \eqref{asymd}, we obtain
\be
|0,in\rangle = {\cal N} e^{\frac{\delta^*}{\gamma}
\int d\tilde p ~a^{out*}_{\tilde p} b^{out*}_{\tilde p}  } |0,out \rangle
\ee
where $\gamma,\delta$ are the Bogolioubov coefficients,
\be
\gamma=\frac{\sqrt{2\pi}}{\Gamma\left(\frac12 + i \frac{M^2}{2\nu} \right)}
e^{-\frac{\pi M^2}{4\nu}+ i \frac{\pi}{2}}\ ,\quad
\delta=e^{-\frac{\pi M^2}{2\nu}} e^{i\frac{ \pi}{2}}
\ee
The vacuum persistence amplitude is therefore given by the overlap
\be
|\langle 0,in | 0, out \rangle|^2 = \exp\left(
-\int d\tilde p ~\ln ( 1 + e^{-\pi M^2/\nu} ) \right)
\ee
This agrees with Schwinger creation rate in two dimensions \eqref{schpart},
upon appropriately interpreting the volume divergence in the
integral over $\tilde p$ \cite{Brout:1990ci}.

\subsection{Physical states at level 0 and 1}
We now determine the low-lying physical states of the bosonic open string in an
electric field, in the framework of old covariant quantization, which is
sufficient for our purposes.
We represent the excited modes in a Fock space built
on a vacuum $|0_{ex}\rangle$ annihilated by the positive
frequency modes $a_{m>0}^\pm$.
Zero-modes in the $(x^+,x^-)$ directions
are quantized using the inverted harmonic oscillator
scattering modes described in the previous section.
In the transverse directions we assume a plane
wave with momentum $a_0^i=k_i$.

\subsubsection{The tachyon}
Let us start with the ground state of the bosonic open string, which
for $\nu=0$ corresponds to a tachyon:
\be
| T\rangle  = \phi(x^+,x^-) |0_{ex} , k \rangle
\ee
The only non-trivial Virasoro constraint is
\be
L_0 | T\rangle = \left[ -\frac12 \left( a_0^+ a_0^- +a_0^- a_0^+ \right)
+ \frac12 \nu^2 -1 + \frac12 k_i^2 \right] | T\rangle = 0
\ee
The tachyon wave function is therefore an eigenmode of the two-dimensional
charged Klein-Gordon equation, with two-dimensional mass
\be
\label{tach}
M^2 \phi :=  \left( a_0^+ a_0^- +a_0^- a_0^+ \right)
\phi= \left(  k_i^2 + \nu^2 -2 \right) \phi
\ee

\subsubsection{The gauge boson}
Now we turn to the first excited level, which for
$\nu=0$ corresponds to a gauge boson.
A general state in this level may be written as
\be
| A\rangle  = \left( -f^+ ~a^-_{-1} - f^-~ a^+_{-1} + f^i ~ a^i_{-1} \right)
|0_{ex} , k \rangle
\ee
where we understood the dependence of $f^{\pm,0}$ on the coordinates $x^+,x^-$.
The mass shell condition $L_0 | A\rangle =0$ requires
\be
[ M^2 - k_i^2 - \nu^2 \mp 2 i \nu] f^\pm = 0 \ ,\quad
[ M^2  - k_i^2 - \nu^2 ] f^i = 0 \ .
\ee
In addition the Virasoro constraint $L_{1}| A\rangle=0$ implies
\be
-(1+i\nu)  a^-_{0} ~f^+ - (1-i\nu) a^+_{0}~ f^-  + a_0^i f^i = 0
\ee
States should be furthermore identified under variation by a spurious state,
\be
\delta | A\rangle = L_{-1} ~ \phi | 0_{ex} \rangle
= \left(- a_{-1}^+ a_0^-  - a_{-1}^- a_0^+  + a_{-1}^i a_0^i \right) ~\phi~
|0_{ex} , k \rangle
\ee
under the condition that the right-hand side still be a physical state.
The $L_0$ and $L_1$ constraints on $\delta | A\rangle$ demands that
\bea
 \left[-(1+i\nu)a_0^-a_0^+ -(1-i\nu)a_0^+a_0^- + a_0^i a_0^i \right] \phi&=&0\\
\left[ M^2  - k_i^2 - \nu^2\right] \phi &=& 0
\eea
These two equations are in fact equivalent, thanks to a 
fortunate conspiration with the value of the
normal ordering constant in \eqref{closedls}.
The spurious state is therefore indeed a physical state,
hence we should therefore identify
\be
(f^+, f^-, f^i) \equiv (f^+ + a_0^+ \phi,\ f^- + a_0^- \phi,\ f^i + a_0^i \phi) \ee
This gauge symmetry may be fixed by choosing
\be
f^+ = a^+_0 \psi,\ f^- = a^-_0 \psi ~~\mbox{with}~~
[ M^2 - k_i^2 - \nu^2  ] \psi = 0
\ee
The $L_1$ constraint allows one to express $\psi$ in terms of the transverse
degrees of freedom, through
\be
-k_i^2 \psi + k_i f^i = 0
\ee
Despite the gap $\nu^2$ in the two-dimensional mass squared,
the first excited level therefore has $D-2$ transverse degrees of freedom,
as required for a massless gauge boson in $D$ dimensions. 
This is satisfying as we do not expect the number of degrees of 
freedom to changed as an electric field background is turned on.
In particular, we see that there are no ghosts up to level 1.

\subsubsection{Closed string states}

Transposing to closed strings in the Milne orbifold, the same
construction can be applied on the left and right movers separately. The
sole effect of the orbifold projection is to restrict the zero-mode wave
function to have integer two-dimensional boost-momentum $\beta J/(2\pi)$, 
determined by the level-matching condition \eqref{matchm}.
In order to enforce this condition it is convenient to
go to a basis where $J$ is diagonal, as we shall proceed to do in Section 4.
For now, we simply notice that this construction
implies the existence of physical states of the bosonic string in each
twisted sector consisting of a scalar tachyon at level (0,0) and
a massless graviton with $(D-2)^2$ transverse degrees of freedom 
at level (1,1), both restricted to having zero boost momentum $j$. 
In addition, there are transverse gauge bosons at level (0,1) and
(1,0), with boost momentum $\pm 2\pi/(w \beta)$. In the type II string,
the tachyon and gauge bosons are projected out, while the Ramond sectors
contribute further states. 

\subsection{One-loop amplitude and Lorentzian physical states}

We now return to Euclidean one-loop
vacuum free energy Eq. \eqref{onelelec}. 
Usually, knowledge of the partition function
determines the spectrum, and indeed this intuition led to the discrete
imaginary spectrum discussed in section 2.2.1. 
Nevertheless, we now show that the same partition
function is consistent with an altogether different spectrum in
Minkowski space. 

To simplify the discussion, we focus on the zero-mode sector, 
where the two prescriptions differ. We may thus restrict
our attention to the factor $1/\sin(\pi \nu t)$ in the Jacobi theta
function in \eqref{jac}, which summarizes the contributions of the
zero-modes, and should be obtainable purely in quantum field theory.
Next we go to a Minkowski
world-sheet, by rotating the Schwinger parameter $t\to it$ in
\eqref{onelelec} (recall that
the expression \eqref{onelelec} assumed
a Lorentzian target space). This converts the zero-mode
contribution into $1/\sinh(\pi\nu\t)$,
which is the expression we now would like to explain.

As is well known (see e.g. \cite{Zuber}, section 4-3-3), the one-loop
amplitude of a charged scalar field in a constant electric field
may be rewritten as the trace of the heat kernel for the
Klein-Gordon operator, \be A_{1-loop} = i \int \frac{ds}{s}
e^{-is(m^2-i\eps)} \mbox{Tr}~e^{i s \Delta_{E}} \ee where
$\Delta_E$ is the electric field Klein-Gordon operator. The main
point of this subsection, and the reason why we have the freedom
to change the quantization scheme in Minkowski space, 
is that the propagator of the charged scalar in an
electric field has two equivalent representations - one as a sum
over discrete states with an imaginary energy and the other as a
sum over a continuum of Minkowski modes. Both are therefore
consistent with the form of the Minkowski one loop partition
function.

The expression as discrete sum over states with imaginary energy
is inherited from the Euclidean continuation
(as already previewed in section 2.2.3, and further discussed in
section 5). We
start with the propagator for the harmonic oscillator $H=-\p_u^2 +
u^2/4$ evaluated in terms of the discrete spectrum, \be \langle
u_1 | e^{-2 i s H} | u_2 \rangle = \sum_{n=0}^{\infty} e^{i
(n+1/2) s} \psi_n^*(u_1)\psi_n(u_2) =\frac{1}{\sqrt{4\pi \sin s}}
\exp\left[ - \frac 14 \left( \frac{u_1^2+u_2^2}{\tan s} - \frac{2
u_1 u_2}{\sin s} \right) \right] \ee The one-loop free energy
in an Euclidean magnetic field
is given by restricting to coinciding points
$u_1=u_2=u$ and integrating over $u$. One obtains:
\be \langle x | e^{-\Delta_{B} s} |  x \rangle =
\int_{-\infty}^{\infty} d\tilde p \langle \sqrt{2/B}(\tilde p+ B x) |
e^{-s H_m} | \sqrt{2/B}(\tilde p+ B x) \rangle = \frac{1}{\sinh(\pi B s)}
\ee  
To go to an electric field in  Minkowski space 
we rotate both $B\to i\nu$ and $s\to i t$, hence reproducing 
the above factor $1/ \sinh(\pi \nu t)$.

However, instead of computing the propagator of the inverted
harmonic oscillator by analytic continuation, the same result may
be obtained directly from the continuous spectrum. Indeed, 
the continuous spectrum
of the inverted harmonic oscillator satisfies the completeness
relation (see e.g. \cite{Moore:1991sf}, eq A 12): 
\bea
\int_{-\infty}^{\infty} \frac{dM^2}{2\nu} &&e^{i \frac{M^2}{2\nu} s }
\psi^\eps(M^2,u_1)\psi^\eps(M^2,u_2)\\
 &=&\frac{1}{\sqrt{4\pi i \sinh s}}
\exp\left[ \frac i4 \left( \frac{u_1^2+u_2^2}{\tanh s} - \frac{2
u_1 u_2}{\sinh s} \right) \right] =\langle u_1 | e^{-2 i s H} |
u_2 \rangle \eea 
valid for $-\pi < \Im(s) < 0$. The same
computation as above therefore leads to the expression $1/
\sinh(\pi\nu s)$, without the need to analytically either $s$ or
$\nu$.

To put it otherwise, 
the contribution of the zero-modes to the one-loop amplitude
may be interpreted either as a sum over the discrete Euclidean spectrum,
or as an integral over the continuous Lorentzian spectrum,
\be
\frac{1}{2i\sin(\nu t/2)}
= \sum_{n=1}^{\infty} 
e^{-i (n+\frac12) \nu t}
= \int dM^2 \rho(M^2) e^{-M^2 t/2}
\ee
where the density of states of the continuous spectrum follows in
the usual manner from the reflection phase shift \eqref{bkw},
\be
\rho(M^2) = \frac{1}{\nu} \log \Lambda
-\frac{1}{2\pi i} \frac{d}{dM^2} \log 
\frac{\Gamma\left(\frac12 + i \frac{M^2}{2\nu}\right)}
{\Gamma\left(\frac12 - i \frac{M^2}{2\nu}\right)}
\ee
where $\Lambda$ is an infrared cut-off.

This argument confirms the validity of our prescription for the
electric case. The one-loop closed string 
amplitude \eqref{milne1l} in the Lorentzian orbifold case does not seem
to be obtainable by a similar
heat-kernel argument. The two systems, however, are markedly
similar and we shall assume that the physical picture we developed
for charged strings carries over to the twisted closed string case
(leaving the physical interpretation of the poles encountered in
the definition of \eqref{milne1l} to future work).

\section{Strings in Rindler space and the Milne orbifold     }
As we have seen, the quantization of the twisted closed string zero-modes
in the Milne orbifold proceeds identically with that of the charged
open string in a constant electric field, only with a further projection
on states with integer boost momentum $\beta J/(2\pi)$. It is thus important
to determine the wave functions in Rindler
coordinates, where the zero-mode boost momentum $j$ is diagonalized and the
orbifold projection most easy to enforce.

\subsection{Open strings in Rindler space coordinates}
Quantization of charged particles in Rindler space was discussed
extensively in \cite{Gabriel:1999yz}\footnote{We became
aware of this work 
after completing a significant part of the analysis. Early references
include \cite{Cooper:1992hw}, and other unfortunate rediscoverers
are \cite{Narozhny:2003ux}.}.
We start by working in the right Rindler wedge, where $X^\pm=\pm r
e^{\pm \eta}/\sqrt{2}$ is a good coordinate system. Since the
zero-mode
boost momentum (or Rindler energy) $j$ commutes with $M^2$, we will
focus on states with a well-defined momentum $j$:
\be f_j(r,\eta)
= e^{-i j \eta} f_j(r)
\ee

\subsubsection{Radial dynamics}

The charged Klein-Gordon equation now
implies the following second order ODE for the radial wave
function,
\be \left[ -r \p_r r \p_r + M^2 r^2 - (j + \frac12 \nu~
r^2 )^2 \right] f_j(r) = 0
\ee
In terms of the coordinate $y=\ln r$,
this can be viewed as the Schr\"odinger equation for a
particle in the potential
\be
\label{rindpot}
V(y) =  M^2 e^{2y} - (j + \frac12\nu ~e^{2y} )^2
\ee
which controls the motion along the radial direction as a function
of Rindler time,
\be
\left( \frac{dr}{r d\eta} \right)^2 + V(r) = 0 \quad \mbox{or} \quad
\left( \frac{dy}{d\eta} \right)^2 + V(y) = 0
\ee
In the absence of an electric field, the
potential \eqref{rindpot} looks like a Liouville barrier, so that all particles
coming from $r=0$ ($y=-\infty$) bounce back at the turning point
$r=M/|j|$. This corresponds to the fact that the classical
trajectories are time-like straight lines in the $(x,t)$ plane, and
therefore reach a finite Rindler radius. For $j=0$ the trajectories
go through the origin.

\FIGURE{ \label{rindtraj}
\hfill\epsfig{file=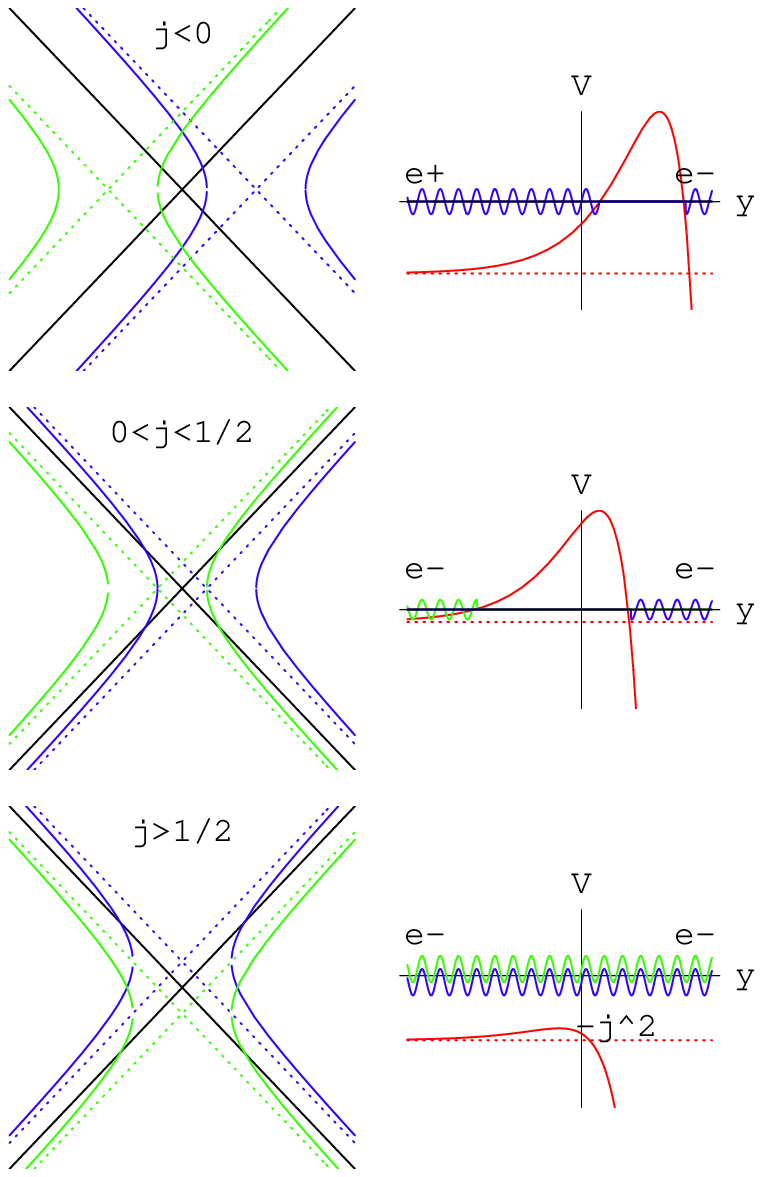,height=19cm}
\hfill
\caption{Classical trajectories in real space and radial dynamics
in the $R$ region. Here $j$ is measured in units of $M^2/\nu$.}
}

For non-vanishing $\nu$, say $\nu>0$,
the situation is more interesting (see Figure 4). At large radius,
the potential is now unbounded from below, corresponding to the
fact that states with $M^2\neq 0$ may now reach $r=\infty$.
If $\nu j/M^2>1$, the potential decreases monotically from $V=-j^2$ at $r=0$
($y=-\infty$). If $1>\nu j/M^2>1/2$, there is
a bump at $r=(M/\nu)\sqrt{2(1-\nu j/M^2)}$, but the energy of the particle
is greater than the height of the barrier $V_{max}=(M^4/\nu^2)
(1-2\nu j/M^2)$.
Finally, if $1/2>\nu j/M^2$, the particle bounces off the barrier,
at the turning points
\be
r_{\pm} = \begin{cases}
\frac{M}{\nu} \left( 1 \pm \sqrt{1 - 2\frac{\nu j}{M^2}} \right) &\ ,\ j>0 \\
\frac{M}{\nu} \left( \pm 1 + \sqrt{1 - 2\frac{\nu j}{M^2}} \right) &\ ,\ j<0
\end{cases}
\ee
These are precisely the extremal values of the Rindler radius $r$
evaluated along the classical trajectory \eqref{traj},
\be
r^2 = -2 X^+ X^- =  -2 x_0^+ x_0^- + \frac{M^2}{\nu^2}
-2\frac{a_0^+ x_0^-}{\nu} e^{\nu \tau/m}
+2\frac{a_0^- x_0^+}{\nu} e^{-\nu \tau/m}
\ee
where we recall that the boost momentum is related to the position of
the center of the classical trajectory through \eqref{boostpart}.
A quantum state of definite $j$ will therefore involve a superposition of many
classical trajectories with a fixed value of 
\be
x_0^+ x_0^-
= \frac{M^2}{\nu^2} \left[ \frac{\nu j}{M^2} -\frac12 \right]
\ee 
Their behaviour can be followed
on Figure \ref{rindtraj}, where we plotted two such trajectories with opposite
values of $(x_0^+,x_0^-)$ for representative values of the dimensionless
parameter $\nu j/M^2$:
\begin{itemize}
\item If  $\nu j/M^2<0$ (Figure 4, top), the electron comes from
$r=\infty$ in the R patch, reaches a minimal Rindler
radius $r_+$, and bounces towards $r=\infty$ 
again. The positron on the other hand comes from the past region $P$,
crosses the past horizon $r=0$ of the $R$ patch $H_R^-$, reaches a maximal
radius $r=r_-$, and falls through the future horizon of the $R$ patch
$H_R^+$ into the future region $F$\footnote{More precisely, the 
wave function describes the time-reverse of this process.}. 
From the point of view of the
radial potential in the $R$ region \eqref{rindpot}, they correspond to
a particle coming from $y=+\infty$ (resp. $y=-\infty$) and bouncing off
the potential barrier back to $y=+\infty$ (resp. $y=-\infty$).
Quantum mechanically, the particle may tunnel under the
barrier and jump from the electron to the positron branch: as
in  static coordinates, this corresponds to stimulated Schwinger emission.

\item If $0< \nu j/M^2< 1/2$ (Figure 4, middle), 
the positron branch now lies entirely
outside the $R$ region. The electron branch of the symmetric hyperbola
centered at $-(x_0^+,x_0^-)$ however does cross the $R$ region, 
entering from the past horizon $H_R^-$ and falling back into the
future horizon $H_R^+$. It corresponds again to a particle coming 
from $y=-\infty$ in the potential $V(y)$, and bouncing off the
barrier.   Quantum mechanically, the particle may again tunnel under the
barrier and jump from the electron branch of one hyperbola to the 
electron  branch of the other: this teleportation process is the analog
of usual Hawking-Unruh pair creation for neutral particles, and proceeds
by nucleation of electron-positron pairs in the middle.

\item Finally, if $\nu j/M^2> 1/2$ (Figure 4, bottom), 
the center of the hyperbola lies
inside the past or forward region. Consequently, all particles either
enter from the past horizon $H_R^-$ and attain future infinity $I_R^+$,
or enter from past infinity $I_R^-$ and fall into the future horizon
$H_R^+$. This corresponds to the fact that the energy of the particle
in the radial potential $V(y)$ is sufficient to cross the barrier.
Even so, part of the wave function will be reflected quantum mechanically,
so that an electron coming from $H_R^-$ may jump to the other hyperbola
and exit the $R$ patch through the horizon $H_R^+$ instead of reaching $I_R^+$.
This process is again the counterpart of Hawking-Unruh pair creation
for neutral particles.

\end{itemize}

From this, we can infer the semi-classical behaviour of the wave functions
in each of the Rindler quadrants, displayed in Figure \ref{rindmodes}.
Notice that the
sign of the charges on each of the asymptotic regions $I_{L,R}^{\pm}$ is
fixed, while the sign of the charges on the horizons $H_{L,R}$ depends
on the sign of $j$.

\FIGURE{ \label{rindmodes}
\hfill\epsfig{file=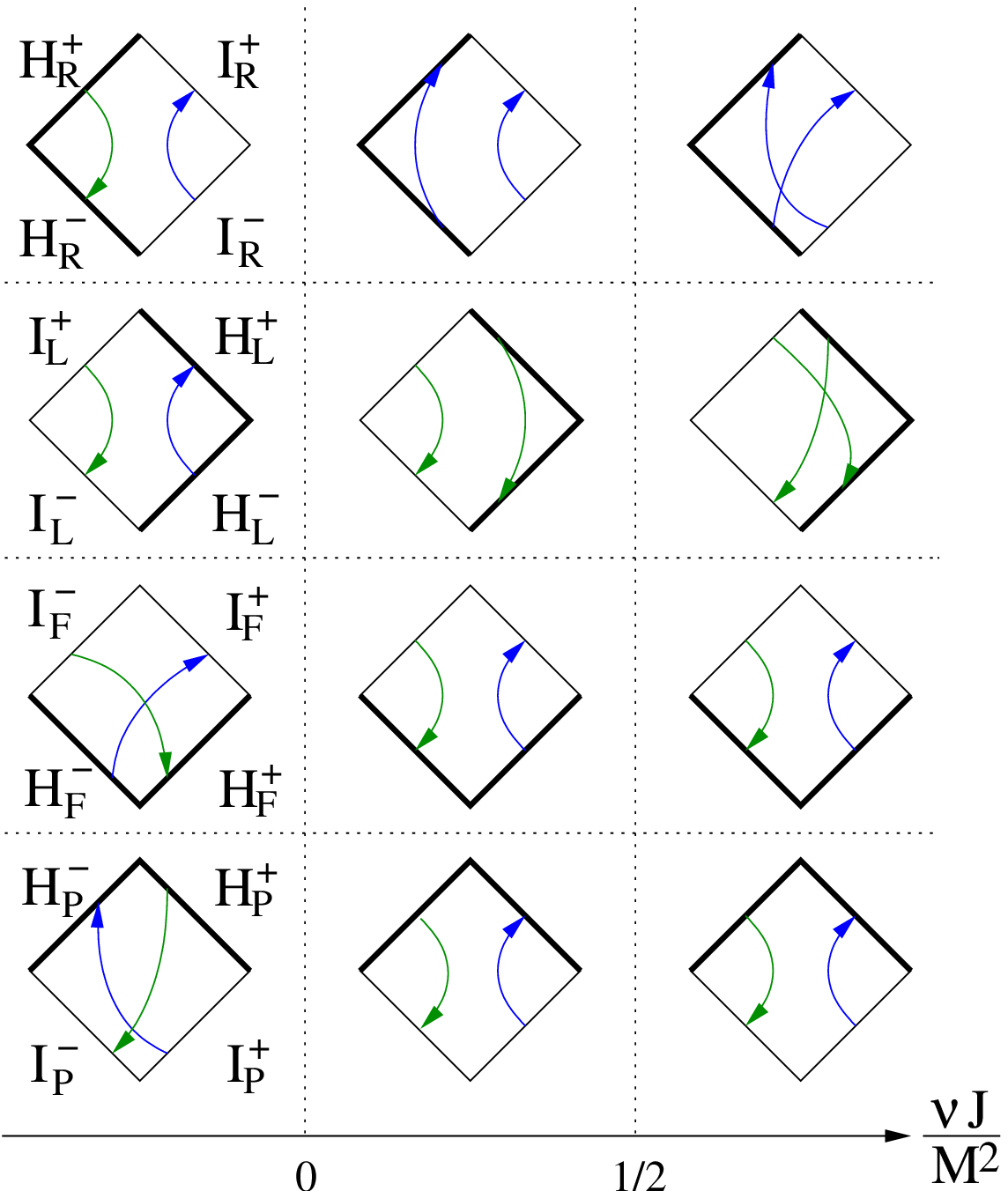,height=12cm}
\hfill
\caption{Classical trajectories in the four
quadrants of Minkowski space.}}

\subsubsection{Normal modes in Rindler wedges}
We may now construct superpositions of wave functions which have only
incoming or outgoing components on either the horizon or the asymptotic 
region, in analogy with the usual modes for a neutral field in Rindler space 
\cite{Fulling:1972md}.
In the R quadrant, incoming modes from $I_R^-$ can be written 
in terms of Whittaker functions as \cite{Gabriel:1999yz} 
\be
{\cal V}_{in,R}^j = e^{-i j \eta} r^{-1}
M_{-i(\frac{j}{2}-\frac{M^2}{2\nu}), -\frac{ij}{2} } ( i \nu r^2 /2 )
\ee
while incoming modes from $H_R^-$ are
\be
{\cal U}_{in,R}^j = e^{-i j \eta} r^{-1}
W_{i(\frac{j}{2}-\frac{M^2}{2\nu}), \frac{ij}{2} } ( -i \nu r^2 /2 )
\ee
The first wave function ${\cal V}_{in,R}^j$ represents an electron
of charge 1 coming in from $I_R^-$, being deflected
by the electric field and exiting at
$I_R^+$; due to pair production, the outgoing charge $q_1$
at $I_R^+$ is greater than the incoming one, and accordingly
there is also an outgoing charge $\sgn(j) q_2$ at the future horizon $H_R^+$.
The transmission and reflection coefficients have been calculated in
\cite{Gabriel:1999yz}, and read\footnote{An absolute value is missing
in \cite{Gabriel:1999yz}, eq (4.14).}
\be
q_1 = e^{-\pi j} \frac{\cosh\left[ \pi \frac{M^2}{2\nu} \right]}
{\cosh\left[ \pi \left( j - \frac{M^2}{2\nu} \right) \right]}\ ,
\quad
q_2 = e^{-\pi \frac{M^2}{2\nu}}
\frac{| \sinh \left[\pi j\right] |}
{\cosh\left[ \pi \left( j - \frac{M^2}{2\nu} \right) \right]}
\ee
Note that these two quantities are positive, and
satisfy $q_1 + \sgn(j) q_2 =1$ by charge conservation.
In contrast, ${\cal U}_{in,R}^j$ describes a particle of charge
$\sgn(j)$ coming in from the past horizon $H_R^-$, and disappearing
into the future horizon $H_R^+$. Due to particle creation, the outgoing
charge at $H_R^+$ is $\sgn(j) q_1$, while there are also $q_2$ electrons
coming out at $I_R^+$. Using these two sets of modes, one can therefore
expand
\be
f(x^+,x^-) = \int_{-\infty}^{\infty}dj
\left[ a^{in}_{{\cal V}_R}(j) {\cal V}_R^j +
\left( \theta(j) a^{in}_{{\cal U}_R}(j) +\theta(-j)
b^{* in}_{{\cal U}_R} (j)
\right)  {\cal U}_R^j \right]
\ee
where $\theta(j)$ is the Heaviside function, and
the oscillators $a$ and $b$ have canonical commutation relations.
Outgoing modes ${\cal U}_{out,R}^j$ and  ${\cal V}_{out,R}^j$ may be
defined similarly by requiring that ${\cal U}_{out,R}^j$ has no component
on $H_R^+$, and  ${\cal V}_{out,R}^j$ has no component on $I_R^+$.
Equivalently, they may be obtained by charge conjugation,
\be
{\cal U}_{out,R}^j (r,\eta) = [ {\cal V}_{in,R}^j (r,-\eta) ]^*\ ,\quad
{\cal V}_{out,R}^j (r,\eta) = [ {\cal U}_{in,R}^j (r,-\eta) ]^*
\ee
Their transmission and reflection coefficients are still given 
in absolute value by $(q_1,q_2)$,
with signs determined by the semi-classical analysis above. The Bogolioubov
transformation relating the incoming and outcoming basis can be found in
\cite{Gabriel:1999yz}. We quote in particular the overlap
\be
| _R \langle 0, in | 0, out \rangle_R |^2
=\exp\left[ - \int_{j<0} dj ~ \ln \left( \frac{1+e^{-\pi M^2/\nu}}
{1+e^{-\pi \frac{M^2}{\nu}+ 2 \pi j}} \right) \right]
\ee
By extracting carefully the volume dependence, the numerator can
be seen to reproduce the standard Schwinger creation rate, while
the denominator is a substraction which scales as
the area, and corresponds to Unruh particle production from the horizon.

The same construction can be carried out in the other wedges. The $L$
wedge is essentially identical to the $R$ wedge, with same reflection
coefficients $(q_1,q_2)$. For the $P$ and $F$ wedges however, the
transmission and reflection coefficients are instead given by
\be
q_3 = e^{\pi \left(j -\frac{M^2}{2\nu} \right) }
\frac{\cosh\left[ \pi \frac{M^2}{2\nu} \right]}
{|\sinh \pi j|}\ ,
\quad
q_4 = e^{-\pi \frac{M^2}{2\nu}}
\frac{\cosh\left[ \pi \left( j - \frac{M^2}{2\nu} \right) \right]}
{|\sinh \pi j|} = q_3 - 1
\ee
In our conventions, $q_3$ and $q_4$ are positive, and occur with signs
determined by the semi-classical analysis in Figure \ref{rindtun}.

\FIGURE{ \label{rindtun}
\hfill\epsfig{file=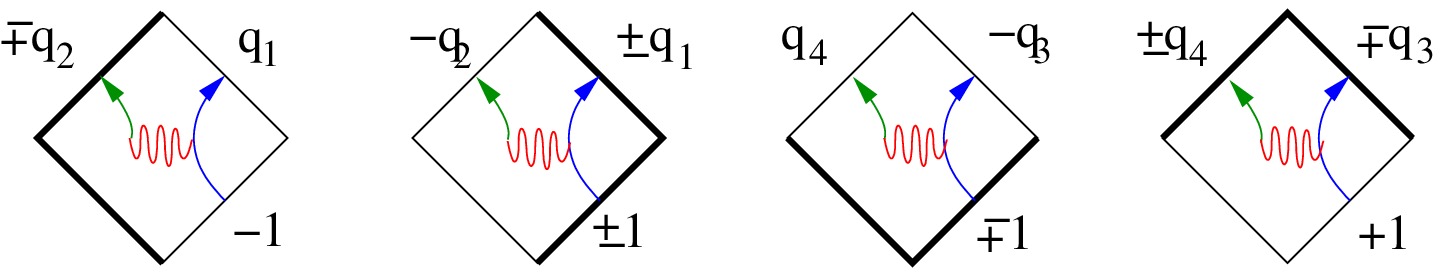,height=2.5cm}
\hfill
\caption{Transmission and reflection coefficients in
the four quadrants of Minkowski space. The lower (upper) sign corresponds
to $j<0$ ($j>0$).}}

\subsubsection{Unruh modes}
Having determined the normal modes on each of the quadrants, it is
a simple matter to fit them into modes on the full Minkowski plane,
without singularities at the horizons: it suffices to start from
the modes in one of the Rindler patches, say $P$,
\bea
\Omega^j_{in,+}&=& {\cal V}_{in,P}^j
= W_{-i(\frac{j}{2}-\frac{m^2}{2\nu}), \frac{ij}{2} } ( -i \nu X^+ X^- )
[X^+/X^-]^{-ij/2}\\
\Omega^j_{in,-}&=& {\cal U}_{in,P}^j
= M_{i(\frac{j}{2}-\frac{m^2}{2\nu}), \frac{ij}{2} } ( i \nu X^+ X^- )
[X^+/X^-]^{-ij/2}
\eea
and analytically continue accross the horizons. The result
can be decomposed in terms of the Rindler modes on each patch,
\bea
\Omega^j_{in,+}&=&
{\cal V}_{in,P}^j  + e^{-i \psi'} \sqrt{q_3}~ {\cal U}_{in,R}^j
- i \sqrt{q_4}~  {\cal V}_{in,L}^j
 - i e^{-i \psi'} \sqrt{q_1}~ {\cal U}_{out,F}^j \\
\Omega^j_{in,-}&=&
{\cal U}_{in,P}^j
+i \sqrt{q_4}~  {\cal U}_{in,R}^j+ e^{i \psi'} \sqrt{q_3}~ {\cal V}_{in,L}^j
 - i e^{-i \psi'} \sqrt{q_1}~ {\cal V}_{out,F}^j
\eea
which create or destroy particles of charge $\pm$ in the past patch $P$
\footnote{The reflection coefficients displayed in
\cite{Gabriel:1999yz}, eq. (6.14) and (6.16) are incorrect.
To ease the notation, the Rindler
modes are tacitly assumed to be truncated to their respective quadrants.}.
In addition, there exists modes that vanish identically in the past,
but can be obtained by analytic continuation of modes in the future
region,
\bea
\omega^j_{in,-}&=& e^{i \psi} \sgn(j) \left[  {\cal V}_{in,R}^j
+ i \sqrt{q_2} e^{i\psi'}  {\cal V}_{in,F}^j \right]\\
\omega^j_{in,+}&=& e^{-i \psi} \sgn(j) \left[  {\cal U}_{in,L}^j
- i \sqrt{q_2} e^{-i\psi'}  {\cal U}_{in,F}^j \right]
\eea
These modes create or destroy particles of charge $\pm$ in the left
or right whisker (see Figure 7). 
In these expressions, the reflection phases are
given by \cite{Gabriel:1999yz}
\be
\psi=\arg\left\{ \frac{\Gamma[ij]}{\Gamma\left[\frac12 +
i \left(j+ \frac{M^2}{2\nu} \right) \right]} \right\}\ ,\quad
\psi'=\arg\left\{
\frac{\Gamma\left[\frac12 + i \left(j-\frac{M^2}{2\nu}     \right) \right]}
{\Gamma\left[\frac12 + i \frac{M^2}{2\nu} \right]}
\right\}\ ,\quad
\ee

\FIGURE{ \label{unruhfig} \hfill\epsfig{file=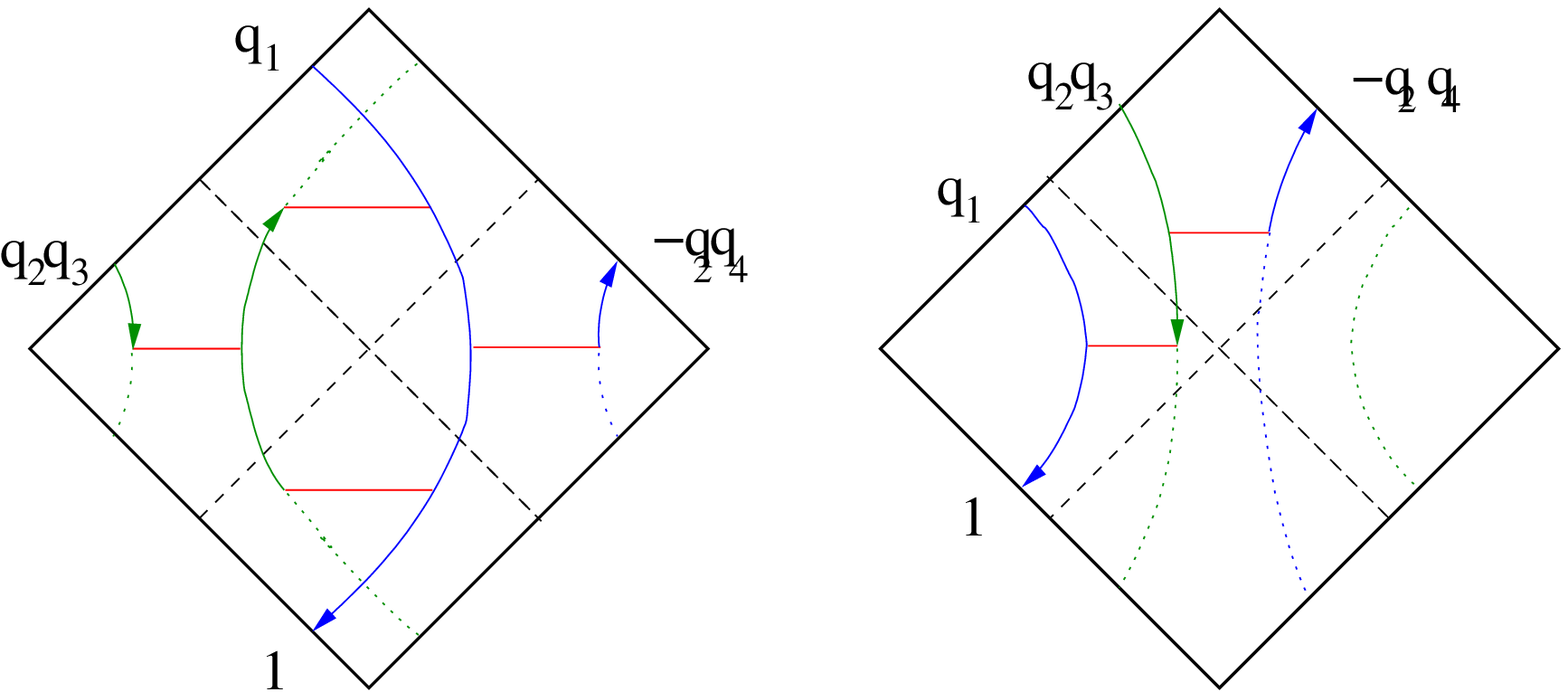,height=6cm}
\hfill \caption{Incoming Unruh modes $\Omega^j_{in,-}$ (left) and  
$\omega^j_{in,-}$ (right). Red lines denote tunnelling events.}}

In the case of the $\Omega_{in,\pm}$ modes, one may see
that charge conservation implies that
$q_1$ particles are produced in the forward region,
of the same type as the incoming particle. This process involves
four tunnelling events.
For the  $\omega^j_{in,\pm}$ modes, which only involves two tunnelling events,
particle creation can only take place
to the future of $L$, and so $\omega^j_{in,\pm}$ vanish on $P$
and on the other side of the whisker. These are the analogue
to the Unruh modes in the neutral case \cite{Unruh:db}. 
Particles of either sign
are produced in the future patch $F$, with charge $\pm(q_2 q_4, -q_2 q_3)$.

\subsection{Particle production in Milne space\label{semirate}}
As we have seen, zero mode wave functions for twisted closed strings
on Milne space in the $w$-th twisted sector are identical 
to those of a charged particle in an electric field $e_0$ such that 
$w\beta=-2\mbox{arcth}(\pi e_0)$, 
upon restricting to integer boost momentum $j$
as specified by the matching condition. Since Schwinger pair production
is equivalent to tunneling under the potential
barrier, it is clear that 
pair production  will take place in the Milne orbifold as well.

In contrast to the electric field however, where pair production occurred
homogeneously throughout space and time, the orbifold projection restricts
the emitted pairs to have fixed integer boost momentum, e.g $j=0$ for
left-right symmetric states such as the tachyon or the graviton.
As represented in Figure \ref{jzero}, classical trajectories of
massive untwisted particles with $j=0$ correspond to straight lines
going from the past region to the future region through the origin;
massless particles on the other hand come from the past to the whisker,
or from the whisker to the future region. For twisted states on the
other hand, classical trajectories with $j=0$
are hyperbolae, one branch of which goes through the origin. One of the member
of the pair therefore goes from the past to the future region,
while the other remains purely in the whisker. Particle production involves
jumping from one branch to the other, hence the particles produced
in the whisker are correlated to those in the future region. From
the point of view of an observer in the whisker (which seems 
to be favored in WZW implementations of the Milne
Universe \cite{Elitzur:2002rt,Elitzur:2002vw}), pair production is
thus described by a density matrix of charged particles of a given
sign.

\FIGURE{ \label{jzero} \hfill\epsfig{file=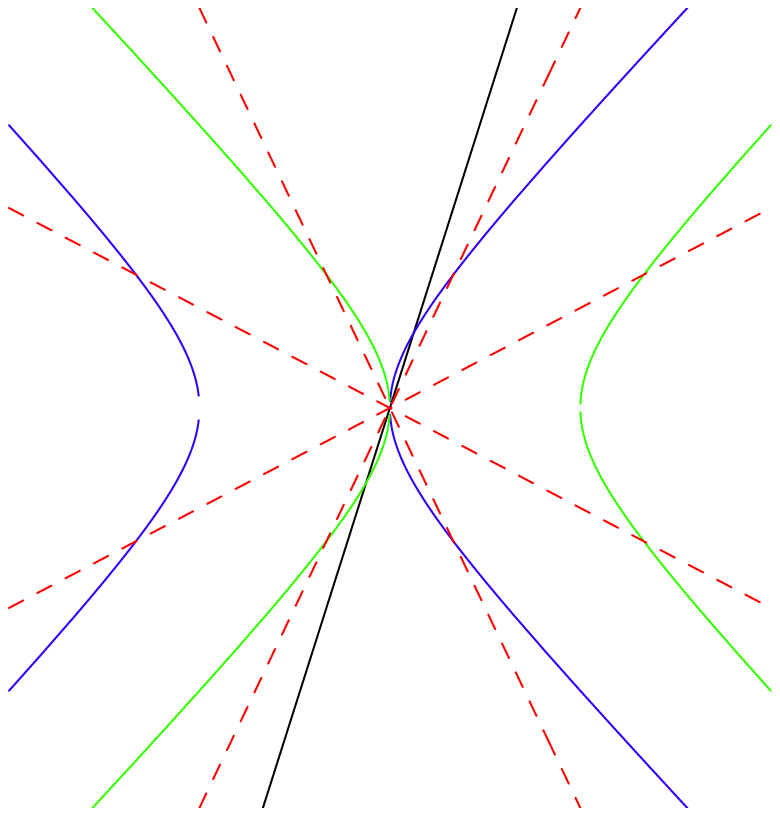,height=5.5cm}
\hfill \caption{Classical untwisted (black) and twisted (blue,green)
trajectories with vanishing boost momentum. In dashed lines, 
a fundamental domain of the Lorentzian
orbifold.}}

One may evaluate the distribution of produced pairs semi-classically,
by assuming an homogeneous distribution in the $(x_0^+,x_0^-)$
phase space (as is the case in the electric field), and asking how
many classical trajectories with a given boost momentum go 
through a given point
$(X^+,X^-)$ in space-time.  The total number of pairs produced
at $(X^+,X^-)$ is therefore given by the
Jacobian 
\be w(X^+,X^-)= \int dx_0^+ dx_0^- \hspace*{1cm}\ee
\be\delta[ (X^+-x_0^+)(X^- -x_0^-)+\frac{M^2}{2\nu^2} ] ~\delta[ \nu
x_0^+ x_0^- +\frac{M^2}{2\nu^2} - j] \nn
\ee 
We find 
\be w(X^+,X^-) =
\frac{1}{\sqrt{(j-\nu X^+ X^-)^2+ 2 M^2 X^+ X^-}}
\ee
For $j=0$ and $M^2>0$, the
production therefore diverges on the light-cone, as well as
around a definite trajectory in the whiskers. For $M^2=0$,
the production diverges at an higher rate on the light cone,
leading to a logarithmically divergent total number of
produced pairs. 

While this copious pair production of twisted states at the singularity
is a promising indication that they may resolve the cosmological singularity,
this is difficult to demonstrate with the current 
string perturbation techniques, as this mechanism involves
condensation of correlated multi-particle states.
Before addressing this challenging backreaction issue, 
an intermediate goal  may be to
send twisted squeezed states from $-\infty$,
along with untwisted states, and see whether
scattering amplitudes are better behaved. While this is still a
daunting problem, a necessary step is to determine the vertex operators
for these twisted states, to which we now turn.



\section{Analytic continuation and vertex operators}
We have constructed above the physical spectrum of open strings in an
electric field, and closed strings in Milne space, in a Lorentz
signature target space. In order to compute
S-matrix elements with more than one incoming particle and one
outgoing particle, or for any genus$>$1 computation, a necessary step
is to analytically continue to Euclidean signature both on
the world-sheet and in target space.
The Euclidean formulation that we will propose consists of two
parts:
\begin{itemize}
\item[(i)] {\it An Euclidean world-sheet/ Euclidean target space
CFT.}
  In the electric field case this is simply a constant magnetic field
in Euclidean space. In the Milne case, the analytic continuation will
turn out to be a rotation orbifold of two-dimensional Euclidean space,
although of a somewhat unusal type: the rotation angle $\theta$ is
equal to the boost parameter $\beta$ of the Milne universe,
and in general irrational. For
$\beta=2\pi p/q\in 2\pi \mathbb{Q}$,
our proposal differs from the usual rotation orbifold by the
fact that twisted sectors with deficit angle $\theta$ and $\theta+2\pi$
should not be identified.
\item[(ii)] {\it A set of vertex operators to be
used for S-Matrix computations in these CFTs.} In the magnetic field
  case, these vertex operators are not the ones associated to the
  states of the Euclidean theory (which are the standard normalizable Landau
  modes of the magnetic field problem), but rather a continuum of non
  normalizable states. In the Euclidean Milne space these are certain
  Wilson line observables on the world-sheet, corresponding to non-normalizable
states in $\Real^2$.
\end{itemize}
Since our derivation of the Euclidean formulation will parallel the
usual analytic continuation from flat Minkowski to Euclidean space,
we start by revisiting the latter, before discussing the electric
field and Milne space in turn.

\subsection{From flat Minkowski space to flat Euclidean space}
We begin with a simple field theory analysis of a scalar field in
flat Minkowski space $\Real^{1,d}$. The path integral describing
the quantum theory is \be \langle {\cal O}(\phi,\phi^*) \rangle
=\int [d\phi] [d\phi^*] ~ {\cal O}(\phi,\phi^*) e^{-i\int dt~ d^3x
{1\over
    2}\eta^{\mu\nu}\partial_{\mu}\phi\partial_{\nu}\phi}
\ee To go to Euclidean target space we rotate the $t$ contour
integral gradually along the family of contours
$t=e^{-i\theta}\tau$ (where $\tau$ runs from $-\infty$ to
$\infty$) from $\theta=0$ (the Minkowski theory) to $\pi/2$ (the
Euclidean theory).

The S-matrix elements of the Minkowskian theory are given by the
residues at the poles in the expression \be \int d^4x_1 \dots d^4 x_n ~
e^{i\Sigma p_\mu^i x^{\mu i}} \langle T[\phi(x_1)....\phi(x_n)] \rangle
\ee
where $p_0^i>0$ for outgoing particles, and $p_0^i<0$ for incoming
particles.  We now rotate the contour integrals $\int dt_k$ inside the path
integral without changing the external momenta. Since we are
rotating dummy integration variables, we remain on the
poles and are still computing the S-matrix although now we are using
an Euclidean path integral. Under this contour
deformation, the vertex operator $\int dt e^{ip_0\tau}$ becomes
a non-normalizable operator $\int d\tau e^{p_0\tau}$
which diverges at $\tau \to \pm\infty$ in order
to allow for the outgoing (incoming) excited state to propagate
to Euclidean time $\tau=+\infty$ (from $\tau=-\infty$).
Of course, these non-normalizable operators need to be regularized,
which can be done by introducing an infrared cut-off and expanding
them on the basis of normalizable states.

Following the same procedure outlined above, the S-matrix
for string theory on the same  flat Minkowski space $\Real^{1,d}$
can be written as the correlation function in the Euclidean
theory,
\be \langle e^{E_1\tau(z_1,{\bar z}_1)+i{\vec p}_1{\vec
x}(z_1,{\bar
    z}_1)}...e^{E_n\tau(z_n,{\bar z}_n)+i{\vec p}_n{\vec x}(z_n,{\bar
    z}_n)}\rangle_E, \ee
integrated over the moduli of the punctured
Riemann surface. Note that the operator $e^{E\tau}$, for real
$E$, does not appear in the state/operator correspondence of the
Euclidean CFT -- the latter are instead $e^{iE\tau}$ for real $E$.
In flat space this does not cause any problem because one can analytically
continue the energies $E_i$ so as to use the
the local operators $e^{iq t}$ of the theory.
In the electric field the analytic
continuation of the Minkowskian wave functions to Euclidean space
is again a class of non-normalizable vertex operators, but now
markedly different than the states of the
Euclidean theory, with no obvious way of continuing from
one class to the other. In appendix B we outline how to
compute the flat space correlators of $e^{E\tau}$ without
analytically continuing $E$.

\subsection{The electric field case}

We are now ready to tackle the electric field case. Our discussion
will be restricted to the field theory limit, since most of the
subtleties are in the zero mode sector. The Lorentzian path
integral in an electric background now is \be \int [d\phi]
[d\phi^*]~ e^{-i\int d^dx dt \left\{ -|(\partial_t-i{\nu\over 2} x
)\phi|^2 +
  |(\partial_x+i{\nu\over 2} t )\phi|^2 \right\}}
\ee
We analytically continue $t$ as before. In order to obtain a
manifestly positive definite expression we will also analytically
continue $\nu$ counter to the rotation of $t$, \ie,
\be t=e^{-i\theta}\tau,\ \ \
\nu=e^{i\theta}b
\ee
where $\theta$ is varied from 0 to $\pi/2$ while keeping $\tau$
and $b$ fixed.  This rotation multiplies the kinetic
term by a phase and does not introduce relative phases between the
derivative and connection terms. The result is, as one might have
expected, a magnetic field $b$ in Euclidean space, described
by the Euclidean path integral
\be \int [d\phi] [d\phi^*]~
e^{-\int d^dx d\tau \left\{ |(\partial_\tau-i{b\over 2}
x )\phi|^2 +
  |(\partial_x+i{\nu\over 2} \tau )\phi|^2 \right\}}
\ee
The same Euclidean theory would have been obtained by
analytic continuation of a magnetic field in Minkowski
space. The
difference between the two cases lies in the choice of vertex
operators to be used in order to compute S-matrix elements
in either theories: to compute the S-matrix in Minkowski
space with a magnetic background one would use vertex operators
which are non-normalizable in the Euclidean time  direction transverse
to the magnetic field, whereas for the electric case at hand
one should use an altogether different set, which are the
analytic continuation of $\phi_{in}^{\pm}$ and $\phi_{out}^\pm$ from
Section 3.2. For instance, the operator $\phi_{\tilde p}^{in,e}$
creating an electron coming from $t=-\infty$
\be
\phi^{in}_e= D_{-\frac12+i\frac{M^2}{2\nu}}
\left[e^{\pi i/4}(\tilde p+\nu x)\sqrt{2/\nu} \right]
 e^{-i(\tilde p + \frac12 \nu x)t}
\ee
rotates to
\be
\phi^{E,in}_e= D_{-\frac12+\frac{M^2}{2b}}
\left[i(\tilde p+ b x)\sqrt{2/b} \right]
 e^{-i(\tilde p + \frac12 b x)\tau}
\ee
where we also rotated the conserved momentum $\tilde p\to i\tilde p$.
This indeed is a non-normalizable
eigenmode of the upward harmonic oscillator with
negative energy $E=-M^2/{2b}$.
Using formula \eqref{asymd} for the asymptotic behavior of the parabolic
cylinder function $D_{-1-p}(ix)$, we see that this indeed blows up at
$x\to +\infty$, where the electron comes from. The same holds for
all the other modes.

It is also interesting to analyze the analytic continuation of the
eigenmodes in Rindler coordinates. Upon rotating $t=-i\tau$, the
light-cone coordinates $X^\pm$ become a pair of complex
coordinates in the two-dimensional plane,
$X^+=(x-i\tau)/\sqrt{2}:=\bar Z, X^-=Z$. The boost momentum $j$
becomes the angular momentum in the $(Z,\bar Z)$ plane, and the
equation for the radial motion in Rindler space just becomes the
Schr\"odinger equation for a two-dimensional harmonic oscillator
in radial coordinates. The Rindler eigenmodes ${\cal V}^j_{in,R}$
and  ${\cal V}^j_{in,R}$ once again become non-normalizable states
of the two-dimensional harmonic oscillator.  ${\cal V}^j_{in,R}$,
which described a particle coming from $I_R^-$, now diverges at
$\infty$ in the complex plane; ${\cal V}^j_{in,R}$, which
described a particle coming from the horizon, now diverges at the
origin of the complex plane. The same issue will arise in the
Milne orbifold case.


\subsection{The Milne orbifold case}

We now make some comments on the continuation of the Lorentzian
orbifold $\Real^{1,1}/boost$ to Euclidean space. We will propose
an analytic continuation but clearly more work is needed to
establish its terms of usage. Under Wick rotation $t=-i \tau$, the
light-cone coordinates $X^\pm$ rotate into a complex coordinate and
its complex conjugate $Z,{\bar Z}$. Wick rotating the boost
parameter $\beta=i\theta$, as in the electric case the orbifold
action now identifies $(Z,{\bar Z})\rightarrow
(e^{i\theta}Z,e^{-i\theta}{\bar Z})$, \ie points on the Euclidean
plane related by a rotation of angle $\theta$. This cannot be the
usual rotation orbifold however, for the following reasons:
\begin{itemize}
\item[i)] The quantum symmetry of the Minkowskian orbifold is
$U(1)$ (acting by multiplying the $w$-th twisted sector by a
character $e^{i\alpha w}$). On the other hand, if $\beta=2\pi
k/N$, then the standard orbifold by a rotation $e^{2\pi i k/N}$
has $N$ twisted sectors and a quantum symmetry $\Zint_N$.
\item[ii)] If $\beta=2\pi r$ for some irrational $r$ (a case that
we will refer to as an ``irrational rotation angle''), we do have
an infinite number of twisted sectors, but the action on the
angular coordinate is ergodic, and smooth untwisted states have to
be rotationally invariant. This disagrees with the integer
quantization of the angular momentum in the Milne orbifold.
\item[iii)] 
Twisted sectors in the usual rotation orbifold describe
normalizable states localized at the fixed points. The Lorentzian
orbifold on the other hand requires non-normalizable vertex
operators with negative Euclidean energy\footnote{In particular,
this seems to rule out the localized tachyon considered 
in \cite{Adams:2001sv}.}.
\end{itemize}
In order to address i) and ii), it is useful to recall that
twisted sectors with irrational rotation angle
have been recently encountered in a covariant treatment of closed strings
in a gravitational wave supported by Neveu-Schwarz
flux \cite{D'Appollonio:2003dr}\footnote{Conical singularities with
an irrational angle also arose in stringy investigations of
black hole and Rindler space thermodynamics
\cite{Dabholkar:1994ai,Lowe:1994ah}, where they were assumed
to reduce to the usual rotation orbifold at
rational values of the deficit angle.}.
This wave background admits a simple description
in terms of a Wess-Zumino-Witten model based on a non semi-simple algebra,
the extended Heisenberg algebra $H_4$ \cite{Nappi:1993ie}.
Using a free field representation,
one may represent the vertex operators of closed string states with
$p^+\neq 0$ by the product of an exponential of the light-cone coordinates
$e^{i (p^+ u + p^- v)}$, times a twist field $H_\theta$ that creates
imposes twisted boundary conditions
$\langle Z(z) H_\theta(0)\rangle \sim z^{\theta/(2\pi)}$
on the transverse coordinates $(Z,\bar Z)$,
with $\theta=2\pi p^+$ \cite{Kiritsis:1994ij}.
Correlation functions for up to four twist fields with arbitrary
level of excitation have been computed using current algebra
techniques \cite{D'Appollonio:2003dr}. A crucial feature in the discussion
is the existence of ``spectrally flowed states'', which appear when the
rotation angle $\theta$ exceeds $2\pi$, thereby lifting the field
identification $\theta\sim \theta+2\pi$ and allowing a smooth
description independent of the rationality of $\theta$ \cite{Kiritsis:1994ij}.
These states have a simple physical interpretation as
long strings winding around the center of the transverse plane,
which are stabilized by a balance
between tensive energy and flux  \cite{Kiritsis:2002kz}.

While this construction does provide a computational framework to
treat twist fields with a continuous rotation angle, it still is
not directly useful for our purposes, as the natural observables
involve normalizable states in transverse space.
In order to deal with non-normalizable wave functions relevant
for the Milne Universe, one may instead consider the
double analytic continuation of the Nappi-Witten background,
\be
ds^2 = -2 dU d\bar U + dX^+ dX^- -\frac14 \mu^2 X^+ X^- dU^2\ ,
\quad H= \mu dU dX^+ dX^-
\ee
where $U$ is a complex coordinate which arises by Wick rotation
of the light-cone coordinates $(u,v)$. Unfortunately, the non-reality
of the metric and flux casts doubts on the consistency of this model.

Given these difficulties, we now suggest an alternative construction of
the irrational rotation orbifold, which is smooth in the
parameter $\beta$, has a $U(1)$ quantum symmetry for any value of
$\beta$ and contains observables for any integer angular
momentum $j$. Our procedure is as follows: we first augment the
Minkowski orbifold by a topological sector on the world-sheet which
is trivial in Minkowski space. After rotating the new
world-sheet theory to Euclidean space, we find that the
topological sector now contains information -- it is the custodian
of the $U(1)$ quantum symmetry even when the rotation angle
is rational. The price to pay is
that vertex operators are no longer
local invariant operators, but involve Wilson lines
on the world-sheet.

In more detail, we replace the usual Lorentzian path integral
over $X^\pm$ by \be \int
[dX^+] [dX^-] [dA] [d\theta]~
e^{i\int \left[ (\partial_\mu X^+ +\beta A_\mu
  X^+)(\partial_\nu X^--\beta A_\nu X^-) \eta^{\mu\nu} + \theta
  F_{01} \right] } \Pi_{i=1}^k \bigl(\Sigma_{n_i}e^{{i2\pi n_i}
  \int_{\gamma_i} A}\bigr) \ee where $\gamma_i,\ i=1 \dots k$, denote the
non-trivial 1-cycles of the Riemann surface, and $\theta$ is a
dynamical field taking values in $\Real$. This action is gauge invariant
under a non-compact Abelian gauge group, \be X^\pm\rightarrow
e^{\pm\beta \alpha}X^\pm,\ \ A\rightarrow A-d\alpha \ee
However, we do not impose any identification on the
coordinates $X^\pm$, but instead use the gauge field to
implemented the twist.
The last factor in the path integral is in fact not
a two-dimensional integral of Lagrangian density, however it can be
absorbed into the latter if we write $\int \theta F$ as $\int
d\theta A$ and assume $\theta$ to be periodic: upon integrating
by parts, one pick up lines in which $\theta$ jumps by its allowed
periodicity, giving rise to the integrals of $A$ over
1-cycles. Notice that this prescription is modular
invariant\footnote{A similar
construction was used in \cite{Giveon:1994fu,
Alvarez:1993qi,Giveon:1993ai,Rocek:1991ps}.}.
The sum $\Sigma_n e^{ {2\pi in}\int_\gamma A}$ serves to restrict
the holonomies around any non-trivial cycle to be integers. Let us
for example consider the torus, represented as usual by
a parallelogram in the complex plane \footnote{Here we are 
using a Euclidean world-sheet and a Minkowski target space.}
with modular parameter $\tau=\tau_1+i\tau_2$. The path integral
splits into a sum over two integers $m_a$ and $m_b$ such that
$\int_{\gamma_a}A=m_a $ and $\int_{\gamma_b}A=m_b$, where
$\gamma_a$ and $\gamma_b$ are the $a$ and $b$ cycles of the torus.
More specifically we can choose a gauge
\be A=d\phi(w,{\bar w}),\ \quad
\phi(w,{\bar w})=m_a~\Re(w)+{m_b-m_a\tau_1\over\tau_2} \Im(w) \ee
The new variables $U^\pm=e^{\pm\beta\phi}X^\pm$ now
have a free field Lagrangian density and are
twisted by $m_a$ and $m_b$ actions of the boost operator along the
two cycles. We then get the standard orbifold partition function
\footnote{In this specific gauge there is still a single
gauge degree of freedom for the entire surface. This fact is
irrelevant here but it will complicate the notion of observables.
We will turn to this point shortly.}.
Finally we rotate to Euclidean target space, $X^+\rightarrow
Z,\ X^-\rightarrow {\bar Z},\ \beta\rightarrow i\theta$. We obtain an
Abelian gauge theory coupled to a unit charge scalar field, with
the only subtlety that the gauge group is non-compact

We can now examine some of the issues raised above. If
$\beta=2\pi/N$ for some $N$, the model we are discussing is still
distinct from the $\Real^2/\Zint_N$ orbifold. The latter would correspond
to a compact $U(1)$ gauge field. In our case the non-compactness
of the Abelian gauge group still enforces selection rules between the
different sector. For example in the pants diagrams it would still
be true that $F=0$ implies $\int A_{in-cycle}=\int
A_{out-cycle-1}+A_{out-cycle-2}$, rather than
this same equality modulo $N$ \footnote{In the
compact $U(1)$ case, the identification
of the holonomy modulo $N$ is implemented by
a large gauge transformation.}.

Let us now consider the observables in the untwisted sector, for
an irrational value of the rotation angle. The standard local
untwisted vertex operators  \cite{Nekrasov:2002kf}
\be \Psi_{p^+,p^-,l}(X^+,X^-)=\int dw~
e^{i(p^-X^+e^{-\beta w}+p^+ X^-e^{\beta w})+ilw} 
\ ,\quad p^+,p^->0
\ee 
satisfy $\Psi_l(e^{\alpha\beta}X^+,e^{-\alpha\beta}X^-)=e^{2\pi
il\alpha}\Psi_l(X^+,X^-)$, and hence are not gauge invariant. This
remains true when we go to the Euclidean
continuation\footnote{Note that the vertex operators are singular
at $Z={\bar Z}=0$.} and replace $X^+$ and $X^-$ by $Z,{\bar Z}$.
However we can still compute S-matrix elements in the following
way. Define an operator \be
\Psi^\gamma_l(Z,\bar Z)= P(e^{2\pi il\int_\gamma A})\Psi_l(Z,\bar Z)
\ee
where $\gamma$ is any path from a fixed reference point, say 0, on the Riemann
surface. Note that this definition is actually independent of the
choice of the trajectory $\gamma$, thanks to the flatness of the gauge
connection $A$ and the quantization law of the holonomies  $m_a$
and $m_b$.
The operator $\Psi^\gamma_l$ is still not gauge invariant -- it
rotates under gauge transformations at point 0. However, if we
have several such operators then the product $\Pi_i
\Psi^{\gamma_i}_{l_i}$ is gauge invariant if $\Sigma l_i=0$. This
is not a new restriction on the model and we expect it to occur in
the orbifold. Hence we can use these operators to define a gauge
invariant $n$-point function. We shall leave twisted vertex operators
for future work.

\section{Discussion}

In this work, we have studied the kinematics
of charged open strings in a constant electric field, and of twisted closed
strings on the Lorentzian orbifold $\Real^{1,1}/boost$. Despite being, to our
knowledge, unrelated under any duality, these two situations share many
formal similarities at the level of first quantized string theory.
In particular, drawing from the well understood dynamics of charged particles
in an electric field, we proposed an alternative quantization
prescription for the open/closed string zero-modes, which does lead to
physical states in the charged/twisted sectors. Despite the lack
of a globally time-like Killing vector, we were also able to give
an analytic continuation to an Euclidean background, in analogy with
the continuation from an electric to a magnetic field. We described the
zero-mode wave functions in a variety of useful representations, and outlined
the corresponding vertex operators, by analytic continuation to the
Euclidean space.
In contrast to the usual observables in a magnetic field or
rotation orbifold, the Euclidean vertex operators describing Lorentzian
scattering states are non-normalizable operators diverging in the
spatial direction where the particle is coming from. They may be expressed
in terms of a continuous spectrum of twist fields in the CFT of a free
complex boson, which remain to be constructed.

While clarifying much of the kinematics, these results are only a first step
toward understanding the dynamics, which we plan to pursue in a future
publication. In particular, the analogy with the electric field falls short
of telling the true fate of the Milne Universe. While twisted state
pair production takes place in both cases, it happens homogeneously
throughout space-time in the electric field case, leading to complete
screening in finite time. In the Milne case, the pair production diverges
on the light-cone as well as in the whiskers, with no effect in principle
at past or future infinity. Yet, the relation between the
electric field $\nu$ and the contraction rate of the orbifold $\beta$
strongly suggests that the Milne singularity may be smoothed under
condensation of twisted states, and possibly lead to a smooth transition
from Big Crunch to Big Bang \cite{Khoury:2001bz}. Indeed, once produced, 
the twisted closed strings contribute an energy that grows linearly with 
the radius of the Milne universe, thus mimicking the effect 
of a two-dimensional positive cosmological constant: the resulting
inflation may thus be sufficient to prevent the circle to reach zero-size.
If so, the non-perturbative
instability toward large black hole formation raised in
\cite{Horowitz:2002mw} may never be reached. A crucial element for
this scenario to hold
is of course that the recombination rate of the twisted strings be lower
than their production rate. Note also that untwisted states may also
be produced, much like uncharged particles in Rindler space. One would
however expect their effect at the singularity to be negligible, as they
become infinitely massive there.

This dynamical slow-down of the contraction rate
may perhaps be understood at the level of low energy
field theory, by going to the T-dual picture.
The collapsing geometry
of Milne turns into a trumpet-like geometry whose circle opens up to
infinite radius in finite time. The string coupling
also blows up there, but the initial coupling (or more
precisely the coupling at some fixed point in the trumpet) may be
made as small as desired such that dilaton gradient is localized
arbitrarily close to $T=0$. Winding modes are now ordinary
momentum modes, and it is not surprising that they should be 
pair-produced. Furthermore, the situation resembles
the standard FRW cosmology of an expanding universe with a gas of particles.
The effect of the latter, for any sensible matter, would be to
decrease the expansion rate of the universe. It may also be
interesting to probe the geometry by localized S-brane probes, although an
early investigation suggested that such probes might see an even
more singular geometry after twisted string
condensation \cite{Nekrasov:2002kf}.

A related question is the fate of the ``whiskers'', which is
tied to the issue of closed time-like curves.
While usually banned on general relativity grounds, they seem
to occur naturally in many string backgrounds, for which they
provide a natural asymptotic region possibly suitable for
holography \cite{Elitzur:2002rt}. Several proposals for shielding
CTC's in string theory have been made in \cite{herdeiro,Boyda:2002ba,
Cornalba:2002nv,Drukker:2003sc}, and it would
be interested to study their fate under winding mode pair creation.
Let us also note that the occurrence of non-normalizable modes
in the Euclidean continuation of the electric field and
Lorentzian orbifold suggest a possible holographic interpretation,
analogous to one of the proposals for holography in plane
waves \cite{Leigh:2002pt}.

Assuming CTC's are still admissible in string theory, the dynamics
in one whisker may after all not be that
complicated. If the analogy to electric field holds, then
positively charged winding states are produced in the
left whisker while negatively
charged winding states go into the right one. Each whisker therefore
sees only one of the two correlated particles, so that
the state in one whisker may be described by a density
matrix of uncorrelated single winding string states.
In string theory this would correspond to
considering the addition of single-particle twisted
states, and at the end sum over all such backgrounds. The
pair creation problem is now transformed into
single particle creation, which could be treated
using CFT techniques, maybe along the lines of \cite{Kazakov:2000pm}.

Irrespective of the details of the dynamics,
it is clear that correlated two-particle states will play a
important role.
This highlights a serious deficiency in
the standard perturbative string theory approach: we do not know how
to take such effects into account systematically.
While condensation of coherent states may be incorporated by the
Fischler-Susskind mechanism, squeezed states and other multi-particle
states require the development of new tools, be it
closed string field theory or non-local string theories,
\cite{Aharony:2001pa,Aharony:2001dp,Berkooz:2002ug,Witten:2001ua,Sever:2002fk}.
These are some of the obstacles that any string theory description of
time-dependent backgrounds will have to address.


\acknowledgments
We are grateful to O. Aharony, C. Bachas, B. Craps, G.
D'Appollonio, J. Distler, B. Durin, S. Elitzur, A. Giveon, E. Kiritsis, A.
Konechny, D. Kutasov, N. Nekrasov, H. Ooguri, E. Rabinovici, I. Schnakenburg
for valuable discussions. M.B. is supported in part by the Israel-US
Binational Science Foundation, the IRF Centers of Excellence program,
the European RTN network HPRN-CT-2000-00122 and the Minerva
Foundation. M.B. would like to thank LPTHE for its hospitality during
early stages of this work.

\appendix

\section{Light-cone quantization in an electric field}
In view of the fact that charged
particles in an electric field are accelerated to the velocity of
the light, it is natural to try and quantize them on the light
front. The world-sheet Hamiltonian \be M^2 = a_0^+ a_0^- + a_0^-
a_0^+ \ee can then be viewed as the generator of dilations in the
$(a_0^+,a_0^-)$ phase space, which allows for very simple
eigenfunctions\footnote{This quantization scheme was first proposed
in a footnote of \cite{Nekrasov:2002kf}, but perhaps too hastily
dismissed.}.
Diagonalizing the generator $P^-=-k_+$ and working
in the $a_0^-=-k_+ - \nu x^-$ representation, we obtain \be
f_{k_+}( x^+, x^-) = (2\nu)^{-\frac14 + \frac{iM^2}{4\nu}}   ~
\Gamma\left(-\frac14 - \frac{iM^2}{4\nu} \right) (k_+ + \nu
x^-)^{-\frac12+\frac{iM^2}{2\nu}} e^{i  x^+ (k_+ + \frac12 \nu
x^-)} \ee This basis of functions is most appropriate to expand
the modes at a fixed (early) $x^-$ time, \ie for incoming
electrons. Equivalently, we may diagonalize the light-cone
momentum $P^+=-k_-$ and work in a  $a_0^+=\nu x^+-k^-$
representation, obtaining: \be f_{k_-}( x^+, x^-) =
(2\nu)^{\frac14 + \frac{iM^2}{4\nu}} ~ \Gamma\left(\frac14 +
\frac{iM^2}{4\nu} \right) (k_- - \nu
x^+)^{-\frac12-\frac{iM^2}{2\nu}} e^{i  x^- (k_- - \frac12 \nu
x^+)} \ee which is appropriate to expand the modes at a fixed
(late) $x^+$ time, \ie for outgoing positrons. The two basis are
related by exchange of position $a_0^+$ and momentum $a_0^-$,
hence by Fourier transform: \be \int \frac{dk_+}{\sqrt{2\pi \nu}}
~ f_{k_-}(\bar x^+,\bar x^-) ~ e^{i k_+ k_- /\nu} =  f_{k_+} \ee
as can be easily checked using the usual 
integral representation of the Gamma function
$A^{-2s} = \frac{\pi^s}{\Gamma(s)}$ $ \int_0^\infty$
$\frac{d\tau}{\tau^{1+s}} e^{-\pi A/\tau}$. The
S-matrix is thus equal to the Fourier transform\footnote{A similar
statement holds true in the context of the $c=1$
string \cite{Alexandrov:2002fh}.}. A full
specification of the characteristic value problem would involve
specifying both the incoming electrons and positrons at
$x^\pm=-\infty$, and reading off the outgoing particles at
$x^\pm=+\infty$, as in \cite{Tomaras:2000ag,Tomaras:2001vs}.

The canonical commutation relations are now easy to determine:
\bea \left[a_{k_+}, a^*_{k_+'}\right] &=& \sgn(k_+ + \nu x^-)
(2\nu)^{-1/2} | \Gamma\left(-\frac14-\frac{iM^2}{4\nu}\right)|^2 
\delta(k_+ - k_+') \\
\left[a_{k_-}, a^*_{k_-'}\right] &=& \sgn(k_- - \nu x^+)
(2\nu)^{-1/2} | \Gamma\left(-\frac14+\frac{iM^2}{4\nu}\right)|^2 
\delta(k_- - k_-') \eea
For positive $k^+ +\nu x^-$, $a^*_{k_+}$ therefore a
creates an electron in the incoming Hilbert space, while for
negative value it annihilates one. Similarly, For negative $k_- -
\nu x^+$, $a^*_{k_-}$ creates an outgoing electron, while for
positive value it annihilates one. These rules have actually a
very simple physical origin:
semi-classically, the modes $f_{k_\pm}$ correspond to charged
particles following the hyperbola \be 2\nu (\bar
x^{\pm}-a'(k_{\pm}) ) ( k_{\pm} \pm \nu \bar x^{\mp} ) = \pm L
\ee where $a(k_\pm)$ is the phase used to construct the wave
packet. For $k^+ +\nu x^->0$, the electron world-line does
intersect the characteristic line at $x^-$, and reaches
$x^+=+\infty$ at a later time. For $k^+ +\nu x^-<0$ however, the
electron has already reached  $x^+=+\infty$ before the initial
time surface; at that time, a new positron has been emitted from
$x^+=-\infty$, and its world-line intersects the characteristic
line at $x^-$.


\section{Correlation function of non-normalizable vertex operators}
Here we briefly discuss how correlation functions of non-normalizable
operators of the form $e^{E\tau}$ may be computed in an Euclidean theory.
One may proceed in two equivalent ways,
\begin{itemize}
\item[(1)] One may regard the insertion of $e^{E\tau(z,{\bar z})}$ at a
  point of world-sheet as imposing the boundary behavior of the fields
  near $(z,{\bar z})$, and compute the Euclidean path
  integral with these boundary conditions.
\item[(2)] One may expand ${\cal O}_E=e^{E\tau}$ as a superposition
  of normalizable modes ${\cal O}_{ip}=e^{ip\tau}$
  (and in principle, descendants as well, although
  that will not be the case here), after introducing an infrared
  cut-off.
\end{itemize}
The motivation for approach (2) is that upon inserting ${\cal
O}_E$ at a point on the world-sheet, the state on a circle of
radius $\nu$ around it is an admissible state in the CFT (this is
also the way to make contact with the first approach), and hence
can be written as sum over operators ${\cal O}_{ip}$ using the
state/operator correspondence of the CFT. One can then use the
standard OPEs to compute the required Green's function. Clearly, a
crucial problem is in enforcing the infrared cut-off in a way
consistent with conformal invariance.

In flat space one may carry out this procedure and verify that it
does give the correct results. The computation of the correlator
$\langle e^{E_1\tau+ip_1x}..e^{E_n\tau+ip_nx} \rangle_E$ proceeds as
follows. Focusing on the $\tau$ CFT we use the regulated expression
above to replace the correlator by \be
\lim_{s\rightarrow 0} \int..\int
dp_1..dp_n \Pi_{i=1}^n \biggl( \int d\tau_i
e^{-s\tau_i^2+E_i\tau_i}\times e^{-ip_i\tau}\biggr) \langle
e^{ip_1\tau(z_1,{\bar z}_1)}...e^{ip_n\tau(z_n,{\bar z}_n)} \rangle_E
\ee The last expression is the familiar $\delta(\Sigma p_i) \Pi
|z_i-z_j|^{p_ip_j}$, and the delta function removes an integration
over momenta. We have to convolute this expression with \be \int
d\tau_i e^{-s\tau_i^2+E\tau_i}\times e^{-ip\tau_i} \propto {1\over\sqrt{s}}
e^{{1\over s}(E_i+ip_i)^2}\ee In the limit $s\rightarrow 0$ we use
steepest descent integration, when we put the $E_i$ on shell and
require $\Sigma_i E_i=0$, to localize the integral to $p_i=-iE_i$. The
end result is the same as the standard analytic
continuation\footnote{It is also easy to track the factors of
  $s$. From $N$ integrations over $\tau_i$, and from $N-1$ integral over
  $p_i$ around the saddle point we obtain a total coefficient of
  $s^{-1/2}$ which is nothing but a single power of the volume, as it
  should be.}.

\section{Parabolic cylinder functions \label{cylpar}}
In this appendix, we assemble some useful formulas involving
parabolic cylinder and Whittaker functions.

We start with parabolic cylinder functions $D_p(u)$, which form a basis of
solutions of the  Schr\"odinger equation for the harmonic oscillator
with an arbitrary (non quantized) energy $E:=p+1/2$:
\be
-\p_u^2+ \left( \frac14 u^2 - E \right) = 0
\ee
Solutions can be expressed as linear combinations of two
among the set
\be
D_p(u), D_p(-u), D_{-1-p}(u), D_{-1-p}(-u)\ .
\ee
Normalizable solutions occur for integer $p$, and can be expressed
in terms of the usual Hermite polynomials,
\be
D_n(z) = 2^{-n/2} e^{-z^2/4} H_n(z/\sqrt{2})
\ee
For general $p$ however, the parabolic cylinder functions are
non-normalizable. As $z\to \infty$ with fixed argument $\theta$,
they admit the asymptotic expansion
\bea
D_p &\sim& e^{-z^2/4} z^ p ( 1+ {\cal O}(z^{-2}) )\ ,\quad
|\theta|<\frac34\pi \\
\label{asymd}
D_p &\sim& e^{-z^2/4} z^ p ( 1+ {\cal O}(z^{-2}) )
-\frac{\sqrt{2\pi}}{\Gamma(-p)}e^{i\pi p} e^{z^2/4}z^{-p-1}
( 1+ {\cal O}(z^{-2}) )\ ,\quad \frac{\pi}{4}<\theta<\frac54\pi \nn\\
D_p &\sim& e^{-z^2/4} z^ p ( 1+ {\cal O}(z^{-2}) )
-\frac{\sqrt{2\pi}}{\Gamma(-p)}e^{-i\pi p} e^{z^2/4}z^{-p-1}
( 1+ {\cal O}(z^{-2}) )\ ,\quad -\frac54{\pi}<\theta<\frac{\pi}{4} \nn
\eea
Eigenmodes of the inverted harmonic oscillator can be obtained by
analytic continuation $u\to e^{i\pi/4}u$, while at the same time
rotating the energy $E\to e^{-i \pi/2} E$, leading to
\be
-\p_u^2+ \left( \frac14 u^2 - E \right) = 0
\ee
Solutions are now linear combinations of any two eigenmodes among
\bea \psi_L^{-}=D_{-\frac12+iE}(e^{-3\pi i/4}u), &&
\psi_L^{+}=D_{-\frac12-iE}(e^{3\pi i/4}u),\\
\psi_R^{-}=D_{-\frac12+iE}(e^{\pi i/4}u), &&
\psi_R^{+}=D_{-\frac12-iE}(e^{-\pi i/4}u)
\eea
corresponding to a purely
outgoing wave toward $u\to -\infty$, incoming wave from  $u\to -\infty$,
incoming wave from $u\to +\infty$ and outgoing wave to $u\to \infty$,
respectively.

More generally, in order to analyze the charged Klein Gordon equation
in Rindler coordinates we are interested in solutions to Whittaker's equation
\be
\label{whittakerode}
\p_z^2 + \left(-\frac14+ \frac{\lambda}{z} + \frac{\frac14-\mu^2}{z^2}
\right) W(z)=0
\ee
Solutions with plane wave behavior at $z\to +\infty$ (resp. $z\to 0$) are
the Whittaker functions
\be
W_{k,\mu}(z) \stackrel{z\to \infty}{\sim} e^{-z/2} z^k\ ,\quad
M_{k,\mu}(z) \stackrel{z\to 0}{\sim} z^{\mu + \frac12}
\ee
These functions are not independent, but satisfy
\be
W_{k,\mu} = \frac{\Gamma(-2\mu)}{\Gamma(\frac12-\mu-k)}M_{k,\mu}(z)
+\frac{\Gamma(2\mu)}{\Gamma(\frac12+\mu-k)}M_{k,-\mu}(z)= W_{k,-\mu}
\ee
For $\mu=-1/4$ we recover the parabolic cylinder functions,
\be
D_p(z) = 2^{\frac14+\frac{p}{2}} W_{\frac14+\frac{p}{2},-\frac14}
\left( \frac{z^2}{2} \right) z^{-1/2}
\ee
Finally, writing $W=e^{-z/2}z^{1/2+\mu}F$
one may rewrite \eqref{whittakerode} as the Kummer equation
\be
\label{kummer}
\left[z \p_z^2 + (1-2\mu - z)\p_z - \left( \frac12 + \mu - k \right) \right]
F = 0
\ee
allowing to express Whittaker functions
in terms of confluent hypergeometric functions,
\bea
M_{k,\mu}(z)&=&z^{\mu +\frac12} e^{-z/2}
\1F1\left(\mu-k+\frac12,2\mu+1,z\right)\\
W_{k,\mu}(z)&=&z^{\mu +\frac12} e^{-z/2}
\2F0\left(\mu-k+\frac12,\frac12-\mu-k,-1/z\right)\\
D_p(z)&=&2^{\frac{p}{2}} e^{-z^2/4}
\left[ \frac{\sqrt{\pi}}{\Gamma\left( \frac{1-p}{2} \right)}
\1F1\left(-\frac{p}{2},\frac12,\frac{z^2}{2}\right)
-\frac{z\sqrt{2\pi}}{\Gamma\left( \frac{-p}{2} \right)}
\1F1\left(\frac{1-p}{2},\frac32,\frac{z^2}{2}\right) \right]
\eea
It is often convenient to use the integral representations,
\bea
\1F1(\alpha,\beta, z) &=& \frac{\Gamma(\beta)}{\Gamma(\alpha)
\Gamma(\beta-\alpha)}
\int_0^1 e^{zt} t^{\alpha-1} (1-t)^{\beta-\alpha-1} dt\\
\1F1(\alpha,\beta,z) &=&\frac{z^{1-\beta}}{B(\alpha,\beta-\alpha)}
\int_0^z e^v ~ v^{\alpha-1} (z-v)^{\beta-\alpha-1} dv \quad
\\
\1F1(\alpha,\beta,z) &=&\frac{2^{1-\beta}e^{\frac12 z}}{B(\alpha,\beta-\alpha)}
\int_{-1}^1 e^{\frac12 z v} ~ (1+v)^{\alpha-1} (1-v)^{\beta-\alpha-1} dv
\eea
where the last two equations are valid for $0<\Re(\alpha)<\Re(\beta)$.


\begin{thebibliography}{00}


\bibitem{Horowitz:ap}
G.~T.~Horowitz and A.~R.~Steif,
``Singular String Solutions With Nonsingular Initial Data,''
Phys.\ Lett.\ B {\bf 258}, 91 (1991).


\bibitem{Khoury:2001bz}
J.~Khoury, B.~A.~Ovrut, N.~Seiberg, P.~J.~Steinhardt and N.~Turok,
``From big crunch to big bang,''
Phys.\ Rev.\ D {\bf 65} (2002) 086007
[arXiv:hep-th/0108187].

\bibitem{Nappi:1992kv}
C.~R.~Nappi and E.~Witten,
``A Closed, expanding universe in string theory,''
Phys.\ Lett.\ B {\bf 293}, 309 (1992)
[arXiv:hep-th/9206078].

\bibitem{Elitzur:2002rt}
S.~Elitzur, A.~Giveon, D.~Kutasov and E.~Rabinovici,
``From big bang to big crunch and beyond,''
JHEP {\bf 0206}, 017 (2002)
[arXiv:hep-th/0204189];


\bibitem{Craps:2002ii}
B.~Craps, D.~Kutasov and G.~Rajesh,
``String propagation in the presence of cosmological singularities,''
JHEP {\bf 0206}, 053 (2002)
[arXiv:hep-th/0205101];

\bibitem{Elitzur:2002vw}
S.~Elitzur, A.~Giveon and E.~Rabinovici,
``Removing singularities,''
JHEP {\bf 0301}, 017 (2003)
[arXiv:hep-th/0212242].


\bibitem{Seiberg:2002hr}
N.~Seiberg,
``From big crunch to big bang - is it possible?,''
arXiv:hep-th/0201039.

\bibitem{Berkooz:2002je}
M.~Berkooz, B.~Craps, D.~Kutasov and G.~Rajesh,
``Comments on cosmological singularities in string theory,''
arXiv:hep-th/0212215.

\bibitem{lms}
H.~Liu, G.~Moore and N.~Seiberg,
``Strings in a time-dependent orbifold,''
JHEP {\bf 0206}, 045 (2002)
[arXiv:hep-th/0204168];
``Strings in time-dependent orbifolds,''
JHEP {\bf 0210}, 031 (2002)
[arXiv:hep-th/0206182].

\bibitem{Balasubramanian:2002ry}
V.~Balasubramanian, S.~F.~Hassan, E.~Keski-Vakkuri and A.~Naqvi,
Phys.\ Rev.\ D {\bf 67} (2003) 026003
[arXiv:hep-th/0202187].

\bibitem{Simon:2002ma}
J.~Simon,
JHEP {\bf 0206} (2002) 001
[arXiv:hep-th/0203201];
J.~Simon,
JHEP {\bf 0210} (2002) 036
[arXiv:hep-th/0208165].

\bibitem{Russo:2003ky}
J.~G.~Russo,
arXiv:hep-th/0305032.


\bibitem{Fabinger:2002kr}
M.~Fabinger and J.~McGreevy,
``On smooth time-dependent orbifolds and null singularities,''
JHEP {\bf 0306}, 042 (2003)
[arXiv:hep-th/0206196].


\bibitem{Horowitz:2002mw}
G.~T.~Horowitz and J.~Polchinski,
``Instability of space-like and null orbifold singularities,''
Phys.\ Rev.\ D {\bf 66}, 103512 (2002)
[arXiv:hep-th/0206228].

\bibitem{bh}
C.~Bachas and C.~Hull,
``Null brane intersections,''
JHEP {\bf 0212}, 035 (2002)
[arXiv:hep-th/0210269];
C.~Bachas,
``Relativistic string in a pulse,''
Annals Phys.\  {\bf 305} (2003) 286
[arXiv:hep-th/0212217].


\bibitem{Fradkin:1985qd}
E.~S.~Fradkin and A.~A.~Tseytlin, ``Nonlinear Electrodynamics From
Quantized Strings,'' Phys.\ Lett.\ B {\bf 163}, 123 (1985).


\bibitem{Abouelsaood:gd}
A.~Abouelsaood, C.~G.~Callan, C.~R.~Nappi and S.~A.~Yost,
``Open Strings In Background Gauge Fields,''
Nucl.\ Phys.\ B {\bf 280}, 599 (1987).


\bibitem{Burgess:1986dw}
C.~P.~Burgess,
``Open String Instability In Background Electric Fields,''
Nucl.\ Phys.\ B {\bf 294}, 427 (1987).

\bibitem{Bachas:bh}
C.~Bachas and M.~Porrati,
``Pair Creation Of Open Strings In An Electric Field,''
Phys.\ Lett.\ B {\bf 296}, 77 (1992)
[arXiv:hep-th/9209032].

\bibitem{Schwinger:nm}
J.~S.~Schwinger,
``On Gauge Invariance And Vacuum Polarization,''
Phys.\ Rev.\  {\bf 82} (1951) 664.

\bibitem{Cooper:kf}
F.~Cooper and E.~Mottola,
``Quantum Back Reaction In Scalar QED As An Initial Value Problem,''
Phys.\ Rev.\ D {\bf 40}, 456 (1989).


\bibitem{Kluger:1991ib}
Y.~Kluger, J.~M.~Eisenberg, B.~Svetitsky, F.~Cooper and E.~Mottola,
``Pair production in a strong electric field,''
Phys.\ Rev.\ Lett.\  {\bf 67} (1991) 2427;
Y.~Kluger, J.~M.~Eisenberg, B.~Svetitsky, F.~Cooper and E.~Mottola,
``Fermion pair production in a strong electric field,''
Phys.\ Rev.\ D {\bf 45}, 4659 (1992);
Y.~Kluger, J.~M.~Eisenberg and B.~Svetitsky,
``Pair production in a strong electric field: An Initial value problem in quantum field theory,''
Int.\ J.\ Mod.\ Phys.\ E {\bf 2}, 333 (1993).



\bibitem{Tomaras:2000ag}
T.~N.~Tomaras, N.~C.~Tsamis and R.~P.~Woodard,
``Back-reaction in light-cone QED,''
Phys.\ Rev.\ D {\bf 62}, 125005 (2000)
[arXiv:hep-ph/0007166];

\bibitem{Tomaras:2001vs}
T.~N.~Tomaras, N.~C.~Tsamis and R.~P.~Woodard,
``Pair creation and axial anomaly in light-cone QED(2),''
JHEP {\bf 0111} (2001) 008
[arXiv:hep-th/0108090].



\bibitem{Bachas:1995kx}
C.~Bachas,
``D-brane dynamics,''
Phys.\ Lett.\ B {\bf 374}, 37 (1996)
[arXiv:hep-th/9511043].

\bibitem{Douglas:1996yp}
M.~R.~Douglas, D.~Kabat, P.~Pouliot and S.~H.~Shenker,
``D-branes and short distances in string theory,''
Nucl.\ Phys.\ B {\bf 485}, 85 (1997)
[arXiv:hep-th/9608024].

\bibitem{Friedmann:2002gx}
T.~Friedmann and H.~Verlinde,
``Schwinger meets Kaluza-Klein,''
arXiv:hep-th/0212163.

\bibitem{Nekrasov:2002kf}
N.~A.~Nekrasov,
``Milne universe, tachyons, and quantum group,''
arXiv:hep-th/0203112.

\bibitem{Cornalba:2002fi}
L.~Cornalba and M.~S.~Costa,
``A New Cosmological Scenario in String Theory,''
Phys.\ Rev.\ D {\bf 66}, 066001 (2002)
[arXiv:hep-th/0203031].



\bibitem{Casher:wy}
A.~Casher, H.~Neuberger and S.~Nussinov,
``Chromoelectric Flux Tube Model Of Particle Production,''
Phys.\ Rev.\ D {\bf 20}, 179 (1979).

\bibitem{Brezin:xf}
E.~Brezin and C.~Itzykson,
``Pair Production In Vacuum By An Alternating Field,''
Phys.\ Rev.\ D {\bf 2} (1970) 1191.

\bibitem{Parikh:1999mf}
M.~K.~Parikh and F.~Wilczek,
``Hawking radiation as tunneling,''
Phys.\ Rev.\ Lett.\  {\bf 85} (2000) 5042
[arXiv:hep-th/9907001].


\bibitem{Dabholkar:1994ai}
A.~Dabholkar, ``Strings on a cone and black hole entropy,'' Nucl.\
Phys.\ B {\bf 439}, 650 (1995) [arXiv:hep-th/9408098].

\bibitem{Lowe:1994ah}
D.~A.~Lowe and A.~Strominger, ``Strings near a Rindler or black
hole horizon,'' Phys.\ Rev.\ D {\bf 51} (1995) 1793
[arXiv:hep-th/9410215].

\bibitem{Kiritsis:1994ij}
E.~Kiritsis, C.~Kounnas and D.~Lust,
``Superstring gravitational wave backgrounds with space-time supersymmetry,''
Phys.\ Lett.\ B {\bf 331}, 321 (1994) [arXiv:hep-th/9404114].


\bibitem{Kiritsis:2002kz}
E.~Kiritsis and B.~Pioline, ``Strings in homogeneous gravitational
waves and null holography,'' JHEP {\bf 0208}, 048 (2002)
[arXiv:hep-th/0204004].



\bibitem{D'Appollonio:2003dr}
G.~D'Appollonio and E.~Kiritsis, ``String interactions in
gravitational wave backgrounds,'' arXiv:hep-th/0305081.



\bibitem{Polchinski:rq}
J.~Polchinski, ``String Theory´´, Vol. 1 and 2, Cambridge Univ.
Press, 1998.



\bibitem{Cornalba:2002nv}
L.~Cornalba, M.~S.~Costa and C.~Kounnas, ``A resolution of the
cosmological singularity with orientifolds,'' Nucl.\ Phys.\ B {\bf
637}, 378 (2002) [arXiv:hep-th/0204261].

\bibitem{Cornalba:2003ze}
L.~Cornalba and M.~S.~Costa, ``On the classical stability of
orientifold cosmologies,'' arXiv:hep-th/0302137.




\bibitem{Maldacena:2000kv}
J.~M.~Maldacena, H.~Ooguri and J.~Son, ``Strings in AdS(3) and the
SL(2,R) WZW model. II: Euclidean black hole,'' J.\ Math.\ Phys.\
{\bf 42}, 2961 (2001) [arXiv:hep-th/0005183].


\bibitem{Moore:1991sf}
G.~W.~Moore, ``Double scaled field theory at c = 1,'' Nucl.\
Phys.\ B {\bf 368} (1992) 557.


\bibitem{Brout:1990ci}
R.~Brout, R.~Parentani and P.~Spindel,
``Thermal Properties Of Pairs Produced By An Electric Field: A Tunneling
Approach,''
Nucl.\ Phys.\ B {\bf 353}, 209 (1991).


\bibitem{Zuber}
C. Itzykson and J.-B. Zuber, "Quantum Field Theory", McGraw-Hill.



\bibitem{Gabriel:1999yz}
C.~Gabriel and P.~Spindel,
``Quantum charged fields in Rindler space,''
Annals Phys.\  {\bf 284} (2000) 263
[arXiv:gr-qc/9912016].


\bibitem{Cooper:1992hw}
F.~Cooper, J.~M.~Eisenberg, Y.~Kluger, E.~Mottola and B.~Svetitsky,
``Particle production in the central rapidity region,''
Phys.\ Rev.\ D {\bf 48}, 190 (1993)
[arXiv:hep-ph/9212206].

\bibitem{Narozhny:2003ux}
N.~B.~Narozhny, V.~D.~Mur and A.~M.~Fedotov,
``Pair creation by homogeneous electric field from the point of view of  an accelerated observer,''
arXiv:hep-th/0304010.

\bibitem{Fulling:1972md}
S.~A.~Fulling,
``Nonuniqueness Of Canonical Field Quantization In Riemannian Space-Time,''
Phys.\ Rev.\ D {\bf 7} (1973) 2850.

\bibitem{Unruh:db}
W.~G.~Unruh,
``Notes On Black Hole Evaporation,''
Phys.\ Rev.\ D {\bf 14} (1976) 870.




\bibitem{Adams:2001sv}
A.~Adams, J.~Polchinski and E.~Silverstein,
``Don't panic! Closed string tachyons in ALE space-times,''
JHEP {\bf 0110} (2001) 029
[arXiv:hep-th/0108075].


\bibitem{Nappi:1993ie}
C.~R.~Nappi and E.~Witten,
``A WZW model based on a nonsemisimple group,''
Phys.\ Rev.\ Lett.\  {\bf 71}, 3751 (1993)
[arXiv:hep-th/9310112].





\bibitem{Giveon:1994fu}
A.~Giveon, M.~Porrati and E.~Rabinovici, ``Target space duality in
string theory,'' Phys.\ Rept.\  {\bf 244}, 77 (1994)
[arXiv:hep-th/9401139].

\bibitem{Alvarez:1993qi}
E.~Alvarez, L.~Alvarez-Gaume, J.~L.~Barbon and Y.~Lozano, ``Some
global aspects of duality in string theory,'' Nucl.\ Phys.\ B {\bf
415}, 71 (1994) [arXiv:hep-th/9309039].

\bibitem{Giveon:1993ai}
A.~Giveon and M.~Rocek, ``On nonAbelian duality,'' Nucl.\ Phys.\ B
{\bf 421}, 173 (1994) [arXiv:hep-th/9308154].


\bibitem{Rocek:1991ps}
M.~Rocek and E.~Verlinde,
``Duality, quotients, and currents,''
Nucl.\ Phys.\ B {\bf 373}, 630 (1992) [arXiv:hep-th/9110053].


\bibitem{herdeiro}
C.~A.~Herdeiro,
``Special properties of five dimensional BPS rotating black holes,''
Nucl.\ Phys.\ B {\bf 582}, 363 (2000)
[arXiv:hep-th/0003063];
C.~A.~Herdeiro,
``Spinning deformations of the D1-D5 system and 
a geometric resolution of  closed timelike curves,''
arXiv:hep-th/0212002;
D.~Brace, C.~A.~Herdeiro and S.~Hirano,
arXiv:hep-th/0307265.


\bibitem{Boyda:2002ba}
E.~K.~Boyda, S.~Ganguli, P.~Horava and U.~Varadarajan,
``Holographic protection of chronology in universes of the Goedel type,''
Phys.\ Rev.\ D {\bf 67}, 106003 (2003) [arXiv:hep-th/0212087].

\bibitem{Drukker:2003sc}
N.~Drukker, B.~Fiol and J.~Simon,
``Goedel's universe in a supertube shroud,''
arXiv:hep-th/0306057.

\bibitem{Leigh:2002pt}
R.~G.~Leigh, K.~Okuyama and M.~Rozali,
``PP-waves and holography,''
Phys.\ Rev.\ D {\bf 66}, 046004 (2002)
[arXiv:hep-th/0204026].

\bibitem{Kazakov:2000pm}
V.~Kazakov, I.~K.~Kostov and D.~Kutasov,
``A matrix model for the two-dimensional black hole,''
Nucl.\ Phys.\ B {\bf 622}, 141 (2002)
[arXiv:hep-th/0101011].

\bibitem{Aharony:2001pa}
O.~Aharony, M.~Berkooz and E.~Silverstein,
``Multiple-trace operators and non-local string theories,''
JHEP {\bf 0108} (2001) 006
[arXiv:hep-th/0105309].

\bibitem{Aharony:2001dp}
O.~Aharony, M.~Berkooz and E.~Silverstein,
``Non-local string theories on $AdS(3) \times  S^3$ 
and stable  non-supersymmetric backgrounds,''
Phys.\ Rev.\ D {\bf 65}, 106007 (2002) [arXiv:hep-th/0112178].

\bibitem{Berkooz:2002ug}
M.~Berkooz, A.~Sever and A.~Shomer,
``Double-trace deformations,
boundary conditions and spacetime  singularities,''
JHEP {\bf 0205} (2002) 034
[arXiv:hep-th/0112264].

\bibitem{Witten:2001ua}
E.~Witten,
``Multi-trace operators, boundary conditions, and AdS/CFT correspondence,''
arXiv:hep-th/0112258.

\bibitem{Sever:2002fk}
A.~Sever and A.~Shomer,
``A note on multi-trace deformations and AdS/CFT,''
JHEP {\bf 0207}, 027 (2002) [arXiv:hep-th/0203168].

\bibitem{Alexandrov:2002fh}
S.~Y.~Alexandrov, V.~A.~Kazakov and I.~K.~Kostov,
``Time-dependent backgrounds of 2D string theory,''
Nucl.\ Phys.\ B {\bf 640} (2002) 119
[arXiv:hep-th/0205079].







\end{thebibliography}
\end{document}